\documentclass{article}

\usepackage[main,final]{neurips_2025}

\usepackage[utf8]{inputenc} % allow utf-8 input
\usepackage[T1]{fontenc}    % use 8-bit T1 fonts
\usepackage{hyperref}       % hyperlinks
\usepackage{url}            % simple URL typesetting
\usepackage{booktabs}       % professional-quality tables
\usepackage{amsfonts}       % blackboard math symbols
\usepackage{nicefrac}       % compact symbols for 1/2, etc.
\usepackage{microtype}      % microtypography
\usepackage{xcolor}         % colors

\usepackage{amsmath}
\usepackage{booktabs}
\usepackage{caption}
\usepackage{subcaption}
\usepackage{graphicx}
\title{PhraseVAE and PhraseLDM: Latent Diffusion for Full-Song Multitrack Symbolic Music Generation}
\author{%
  Longshen Ou \\
  School of Computing\\
  National University of Singapore\\
  \texttt{oulongshen@u.nus.edu} \\
  \And
  Ye Wang \\
  School of Computing\\
  National University of Singapore \\
  \texttt{wangye@comp.nus.edu.sg} \\
}

\begin{document}
\maketitle
\begin{abstract}

This technical report presents a new paradigm for full-song symbolic music generation. Existing symbolic models operate on note-attribute tokens and suffer from extremely long sequences, limited context length, and weak support for long-range structure. We address these issues by introducing PhraseVAE and PhraseLDM, the first latent diffusion framework designed for full-song multitrack symbolic music. PhraseVAE compresses an arbitrary variable-length polyphonic note sequence into a single compact 64-dimensional phrase-level latent representation with high reconstruction fidelity, allowing a well-structured latent space and efficient generative modeling. Built on this latent space, PhraseLDM generates an entire multi-track song in a single pass without any autoregressive components. The system eliminates bar-wise sequential modeling, supports up to 128 bars of music (8 minutes in 64 bpm), and produces complete songs with coherent local texture, idiomatic instrument patterns, and clear global structure. With only 45M parameters, our framework generates a full song within seconds while maintaining competitive musical quality and generation diversity. Together, these results show that phrase-level latent diffusion provides an effective and scalable solution to long-sequence modeling in symbolic music generation. We hope this work encourages future symbolic music research to move beyond note-attribute tokens and to consider phrase-level units as a more effective and musically meaningful modeling target.\footnote{Demos and code: \url{https://www.oulongshen.xyz/midi_ldm}.}

\end{abstract}

\section{Introduction}

% The song is the core unit for all music activities.
The significance of the song is paramount\footnote{We use the word \textit{songs} to indicate \textit{complete pieces of music composition}, not necessarily related to specific forms or genres of music.}. When engaging with music, one interacts with complete compositions or pieces, not fragments. In musical performance, music is executed at the song level. For educational purposes, music is learned through individual songs. Composers and arrangers create music in the form of songs. Artists market their compositions as songs and collections of songs. The rationale is straightforward: analogous to how written content is structured into articles and visual media into complete films or television episodes, the song represents the most prevalent form of auditory art that offers the audience a full, cohesive musical experience.

% So full song generation is an important problem for Music AI
Despite the centrality of songs in musical practice, most music generation systems still operate at the segment level, producing short excerpts rather than complete musical works \cite{wang2024whole}. This disconnect arises not from a lack of demand, but from the inherent difficulty of modeling long-form symbolic music. A full song typically spans tens or hundreds of bars, involves multiple sections with distinct structural functions, and expresses musical ideas through long-range development rather than isolated moments. Consequently, generating a song requires capturing not only local note-to-note coherence but also global structure, such as sectional form, thematic recurrence, harmonic pacing, texture arrangement, and long-term melodic evolution. Bridging this gap between short-form generative models and real-world musical works motivates the need for more effective full-song symbolic music generation.

% A bit details about challenges of full-song modeling for current models
Modeling music is an important topic in generative AI, yet full-song symbolic music generation remains relatively underexplored. Existing approaches typically rely on sub-note-level tokenization, which we call \textit{note-attribute token} in the paper—where a single note event is decomposed into multiple tokens—and autoregressive decoding, such as \cite{huang2020pop,von2023figaro,ouunifying,wang2025notagen,lu2023musecoco}. This design leads to extremely long token sequences, which introduces several inherent limitations. First, the computational cost becomes prohibitive: for instance, generating a 32-bar sample with MuseCoco \cite{lu2023musecoco} can take several minutes, whereas typical users may expect results within seconds. Second, many systems support only segment-level generation because their supported context length are insufficient to model an entire song \cite{ouunifying,zhao2023accomontage,yang2025meteor}. Third, the excessive sequence length makes it difficult to learn long-range musical structure, such as maintaining coherent melodic variation across repeated sections, especially under limited training data \cite{zhao2023accomontage}.

% Related works on full song generation are problematic
Several recent works have begun exploring full-song symbolic music generation \cite{wang2024whole,chen2025segment}. These studies represent important early steps toward modeling long-form musical structure. However, they also face two notable limitations that make them difficult to extend to the multitrack setting considered in this report. First, the generation process in these frameworks is restricted to single-track output; extending their formulations to handle arrangements or multiple instrument voices is non-trivial and would require substantial architectural modification. Second, these methods rely on song-level autoregression, where short segments must be generated one after another. 
This sequential dependency not only results in slow inference, but also imposes a strong inductive bias on how global musical structure is formed, which may limit the model’s ability to plan sections holistically (as further discussed in \S~\ref{sec:motivation_diffusion}). 

% So we present something cool.
To address the challenges discussed above, we present a new hierarchical framework for full-song symbolic music generation. The system removes segment-wise autoregressive generation so that it has high inference efficiency. It can be trained on a relatively small dataset, and its memory footprint is low enough that the entire 800-song corpus can be loaded in a few batches. The framework operates in a compact 64-dimensional phrase-level latent space, where each latent vector represents both the note content and timbre of one bar for a single instrument. 
% While the latent diffusion model alone is sufficient for unconditional generation, placing additional length and structural conditions further improves its usability and long-range coherence.

\begin{figure}[tb]
     \centering
     \includegraphics[width=1.0\textwidth]{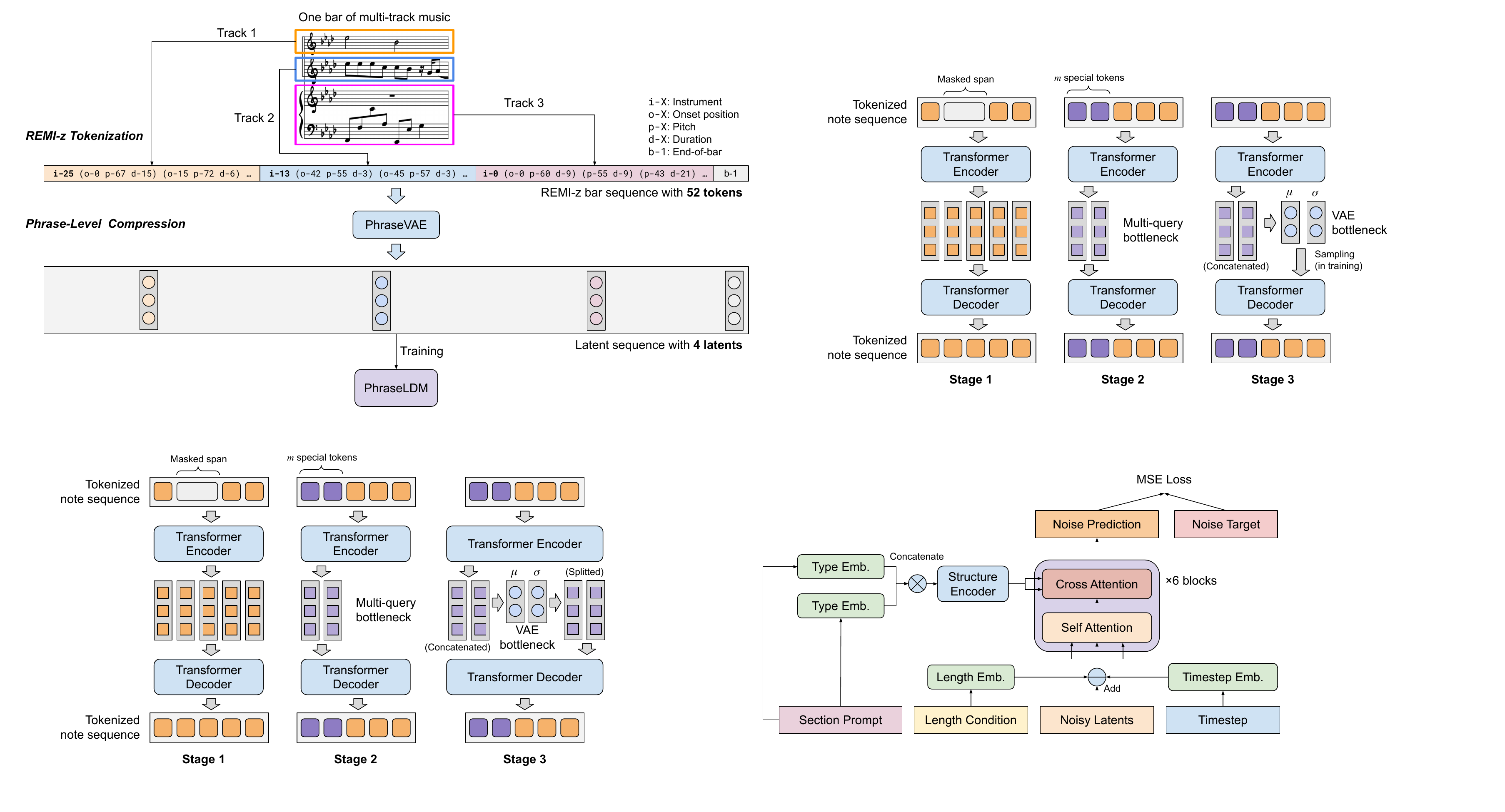}
     \caption{Data representation and framework overview.}
     \label{fig:overview}
\end{figure}

Our main contributions are as follows:

\begin{itemize}

\item \textbf{A Phrase-Based Generation Paradigm.} We propose generating multitrack symbolic music at the phrase level—musically meaningful spans of note sequences—rather than using note-attribute tokens. This paradigm reduces the context length required by the full-song generation model from over 10k tokens to 512 latents, making full-song generation feasible.

% \item \textbf{PhraseVAE.} We introduce a phrase-level sequence compression model that maps variable-length polyphonic note sequences, together with their instrument identities, into compact 64-dimensional latent vectors with near-perfect reconstruction quality (99.0\% F1$_{op}$). PhraseVAE adopts a multi-query compression mechanism that outperforms both query-based and pooling-based baselines, and its multi-stage training strategy accelerates convergence while producing a well-regularized latent space.

\item \textbf{PhraseVAE.} We introduce PhraseVAE, a phrase-level sequence compression model that maps variable-length polyphonic note sequences, together with their instrument identities, into compact 64-dimensional latent vectors with near-perfect reconstruction quality (99.0\% F1$_{op}$). 
Technically, PhraseVAE contributes two key innovations:  
(1) a \textit{multi-query compression} mechanism that enables high-fidelity latent representation learning and consistently outperforms both pooling-based and single-query baselines; and  
(2) a \textit{progressive bottleneck} training strategy that tightens the latent space in stages, yielding compact yet well-regularized representations.  
% These design principles are not specific to symbolic music and may be broadly useful for sequence-level representation learning in other domains.

\item \textbf{PhraseLDM.} 
Built upon the latent space provided by PhraseVAE, we develop PhraseLDM, a latent diffusion model capable of generating an entire multitrack song in a single pass without any autoregressive components. By operating directly on phrase-level latents, PhraseLDM produces coherent local texture and harmony while exhibiting non-trivial structure planning ability. It achieves strong unconditional generation quality, and its controllability is further enhanced through optional conditioning on sequence length and structural layout.

\item \textbf{A Lightweight Full-Song Generation Framework.} Together, PhraseVAE and PhraseLDM form a lightweight yet capable full-song symbolic generation framework. The system contains around 45M parameters in total and can generate a complete multi-track song (up to 128 bars; roughly eight minutes at 64 BPM) within 3.3 seconds on an NVIDIA A40 GPU. This design enables fast, scalable, and practical full-song generation, supporting long-form multi-track compositions with realistic instrument usage and clear global structure.

\end{itemize}

\section{Motivation}

\subsection{Semantics of Music}

% In natural language, word is expressive
In natural language, a single word can convey rich and highly abstract meaning. A word functions as a compact pointer to real-world concepts, such as objects, actions, emotions, or events. For example, the word “encore” immediately evokes a complete situational context: an enthusiastic audience, repeated applause, the anticipation of an additional performance, and the emotional atmosphere surrounding it. In other words, a single lexical token can trigger a complex semantic representation in the human mind. This semantic density is one of the reasons why modern NLP builds on learning expressive representations for words and subwords, enabling models to encode the nuanced meaning they carry \cite{mikolov2013efficient,devlin2019bert}.

% In symbolic music, note-attribute token is not the same
In symbolic music, this semantic richness does not exist at the note or sub-note level. Most symbolic music models—including nearly all autoregressive ones—encode each note event by decomposing it into multiple note-attribute tokens, where each token represents only a single aspect of the event \cite{huang2020pop,qu2024mupt,wang2025notagen,wu2024melodyt5,pasquier2025midi,ouunifying,von2023figaro}. For example, in the REMI-z tokenization scheme \cite{ouunifying}, a token such as \texttt{o-12} merely indicates that an onset occurs at the second beat of a bar. It provides no information about the pitch, instrument, timbre, duration, expressive intent, or the musical role of the note (e.g., whether it belongs to a melodic line, comping pattern, padding texture, or arpeggio). Ironically, our previous model in \cite{ouunifying} learns a 768-dimensional embedding for that token whose semantic content is simply ``something begins here.'' While the efficiency issues of such representations will be discussed later, the key point is that an individual note-attribute token carries almost no semantic meaning on its own, making it a poor unit for modeling music.

% What is expressive is note sequence, i.e., phrase
We hypothesize that the meaning of music is carried by note sequences, not by single note-attribute tokens. A note sequence can involve multiple instruments: when all the strings play together, it often gives a solemn feeling; when the drum fills the empty space before a new section, it suggests the music is moving forward. Even a single-track sequence can be expressive. A loose guitar strumming pattern can sound relaxed and casual, and a fast upward bass line can feel like the player is showing off. A solo banjo with an alternating-bass pattern can create a countryside atmosphere, while a slow free-style erhu line often suggests sadness or mourning. Listeners may not agree on the exact meaning of each example, but most would agree that these sequences do convey something or trigger some emotion \cite{huron2006sweet,temperley2004cognition}. This is where the semantic meaning of music appears. We believe these are the things that need to be captured in composition and in generative music models.

\subsection{Previous Paradigms}

There are two major paradigms of symbolic music generation.

\subsubsection{Autoregressive Models}

% Autoregressive models face the long sequence problem.
The first and most widely studied paradigm is autoregressive modeling. Many works follow this direction \cite{Huang2018MusicTG,wu2024melodyt5,ren2020popmag,huang2020pop,qu2024mupt,wang2025notagen,ouunifying,pasquier2025midi,jiang2025versatile,guo2025moonbeam,thickstun2023anticipatory,lu2023musecoco,von2023figaro}. Some explore large-scale training \cite{qu2024mupt,lu2023musecoco,wang2025notagen}, some use attribute-based conditioning \cite{lu2023musecoco}, some adapt instrumentation \cite{ouunifying,yang2025meteor}, and some incorporate human preference into the training loop \cite{wang2025notagen}. These are all valuable studies, and many of them can generate high quality music segments. However, listeners do not usually consume music in segments. Generating a full piece of music remains a major challenge for most autoregressive models.

To illustrate the difficulty of full-song generation problem, consider a typical multitrack dataset such as Slakh2100. A song in this dataset contains about 10,874 notes at the 95th percentile. Even with an efficient tokenization scheme that converts each note into only three attribute tokens, the sequence still exceeds 30k tokens—far longer than the context length supported by most existing symbolic music generation models. In practice, many autoregressive models are trained only on short segments instead \cite{lu2023musecoco,ouunifying}, and they are simply not designed for full-song generation.

% There are trials, but not good enough
Many previous works have discussed this long sequence issue. Several attempts have been made to reduce sequence length \cite{Hsiao2021CompoundWT,fradet-etal-2023-byte,ouunifying}, but the problem is still not fully solved. The most efficient tokenization schemes \cite{Hsiao2021CompoundWT} use a single compound vector to represent a note event, but even this may not be efficient enough to support full-song generation. This raises a natural question: is there a way to represent note sequences, which we believe are the basic units of composition, at a more abstract level while still preserving enough detail? If such a method exists, it could solve both the efficiency and context length issues, and it may also make long-range modeling easier, such as capturing relationships between different sections of a song.

\subsubsection{Diffusion Models}
\label{sec:motivation_diffusion}

The second paradigm is the use of diffusion models, with a focus on discrete diffusion models \cite{min2023polyffusion,wang2024whole,yuan2025diffusion} in recent years. In this setting, diffusion is applied directly in the data space of the piano roll, which is a matrix-like representation of short single-track music segment. The piano roll usually covers one or a few bars along the time axis and uses the pitch axis as the other dimension, where each cell stores duration and possibly velocity. The model generates one segment at a time. After finishing the diffusion process for that segment, it moves on to the next one, and continues in this way until the song is complete.

This paradigm has several issues. First, it is slow: each segment must go through a full diffusion process, and a full song requires repeating this many times. Second, the data dimension is very high. With a 48th-note resolution, one bar contains 6,144 channels of information in float numbers that the diffusion model must handle (in comparison, ours uses 152.3, as shown later). Third, current attempts are limited to single-track music, and it is not clear how this design can be extended to multitrack composition. Fourth, bar-wise sequential modeling is a questionable inductive bias. We argue that similar bars or sections in music are mutually related rather than strictly dependent in one direction. Even in human composition, composers and arrangers do not always create music in a monotonic left-to-right manner. 

There is also an early attempt at applying latent diffusion to symbolic music \cite{mittal2021symbolic}. It is a valuable demonstration that the LDM framework is, in principle, applicable to symbolic music modeling. However, many issues remain unresolved. 

(1) \textbf{Insufficient VAE reconstruction quality.} The VAE reports only around 90\% token-level accuracy under teacher forcing and around 80\% accuracy when reconstructing directly from latents. This level of fidelity is far from enough to produce natural music with idiomatic playing patterns. A later VAE work \cite{wang2020pianotree} improves reconstruction quality, but we will show later that it does not handle 48th-note quantization well—a temporal resolution widely adopted in symbolic generation tasks.

(2) \textbf{Restricted to monophonic sequences.} The model is designed for monophonic note sequences only. It cannot represent complex polyphonic instruments such as piano or guitar, and the method does not generalize to multitrack music in a straightforward manner.

(3) \textbf{Limited long-term modeling support.} It supports sequences up to 32 bars, which is insufficient for most contemporary full-song compositions.

(4) \textbf{Limitations in music quality.} The LDM outputs are musically weak even for short monophonic segments. The generated samples lack naturalness, coherence, musicality, and structural organization. Overall, while the work is an early step toward applying LDM to symbolic music, the resulting music is far from sounding like real compositions, leaving substantial space for improvement.

Another recent work \cite{li2025erld} has also explored latent diffusion models for symbolic music generation, with a primary focus on harmony-constrained generation. In contrast to this line of work, our study targets full-song multitrack symbolic music generation and places particular emphasis on high-fidelity phrase-level compression, scalable latent representations, and long-range structural modeling across entire compositions.

\section{Method}

% Summary of Methodology
As discussed earlier, full-song generation becomes difficult when the model operates directly on note-attribute tokens, since each token encodes only a small part of a note and leads to extremely long sequences. Our approach is to compress music into a higher-level granularity that carries meaningful musical semantics, which greatly reduces the sequence length required for full-song modeling. The generation model then operates on this compact and abstract representation rather than on raw tokens.

\subsection{Phrase and REMI-z Grammar}

% Phrase’s definition
We now formally define the meaning of a \textit{phrase} used in this paper. A phrase can have different meanings in different music research settings, but here it has a precise definition. A phrase refers to all information of what one instrument plays within a single bar. A phrase must contain the following: the instrument identity, and the set of notes played by that instrument in that bar. Each note includes its onset and duration, as well as its pitch. This is the minimum information required for a phrase, and it is also the level of representation that our generation model operates on.

% How a musical phrase connects with REMI-z
When we prepare symbolic music using the REMI-z tokenization scheme \cite{ouunifying}, the concept of a phrase directly corresponds to the \textit{track sequence} defined in the REMI-z grammar. A phrase (or track sequence) contains the following tokens: it begins with an instrument token, followed by a sequence of note-related tokens. Each note is represented by three parts: an onset token that marks the relative starting position within the bar, a duration token, and a pitch token represented as an integer between 1 and 128. Both onset and duration are quantized with 48-note time resolution. The order of notes follows the chronological order of their onsets. Notes sharing the same onset position use only a single onset token, and their relative order is sorted by pitch, with higher pitch appearing earlier.

% How phrases form a song
A bar contains multiple phrases, one for each instrument that plays in that bar. In REMI-z tokenization, a bar sequence is simply a list of phrase, and the order of the phrases is determined by the average pitch of each corresponding phrase, phrases with higher pitches appears earlier. A song is then formed by a stream of bar sequences. Viewed this way, compressing music at the phrase level is intuitive and greatly reduces the complexity of the information that the generation model needs to handle.

\subsection{PhraseVAE}

% Why phrase is a sweet spot for music representation
We look for a unit of representation that carries meaningful semantics, is relatively independent from its surrounding context, interacts with other units to form musical patterns, and is not so complex that it requires a very large latent dimension to encode. We find that phrase-level compression fits this purpose well. Our goal is to learn a vector representation for a sequence of note events that contains all the essential information of that phrase: the instrument identity, the onset times, the pitches, and the durations. The method must also handle sequences of different lengths, since the number of notes played by different instruments, or even by the same instrument across different bars, can vary widely.

\begin{figure}[tb]
     \centering
     \includegraphics[width=0.95\textwidth]{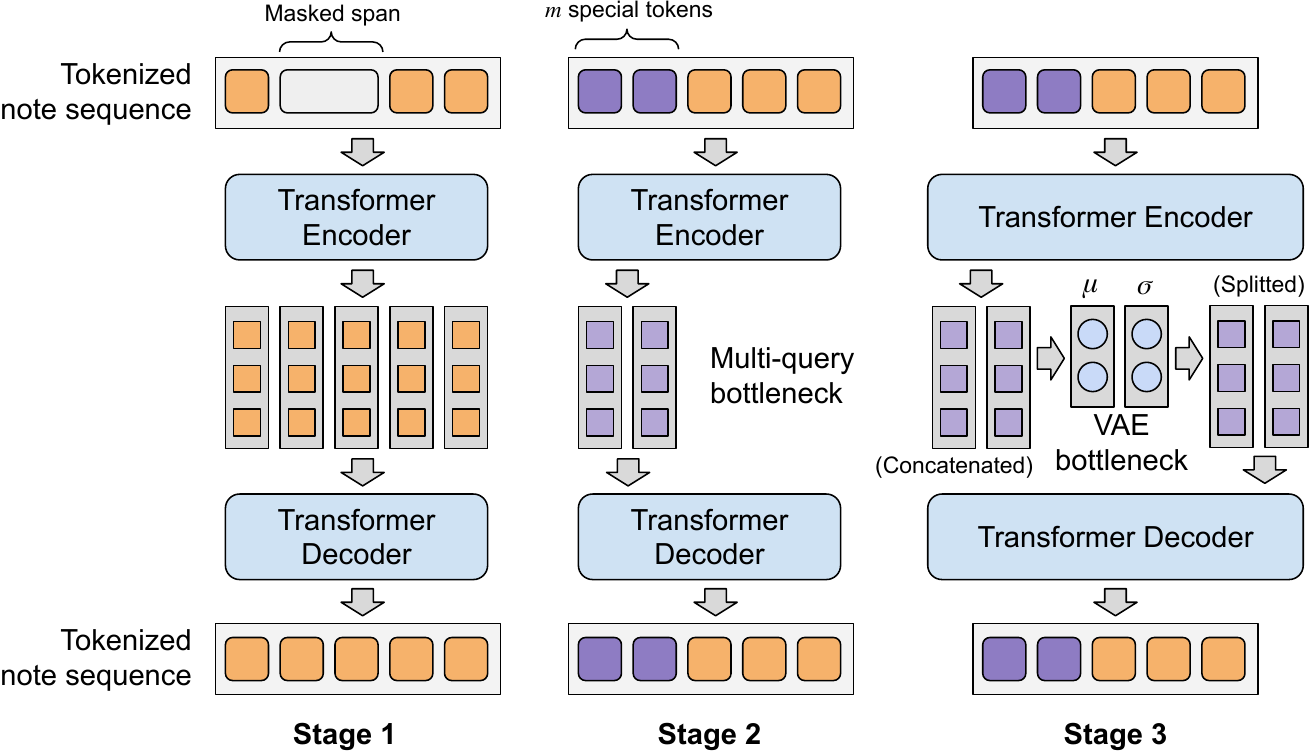}
     \caption{PhraseVAE training stages.}
     \label{fig:vae}
\end{figure}

% Summary of PhraseVAE training workflow
Based on this idea, we build a variational auto-encoder (VAE) that compresses a variable-length note sequence into a 64-dimensional vector with high reconstruction quality (99.0\% F1 when measuring onset and pitch). As in Figure~\ref{fig:vae}, the training procedure follows a three-step recipe. First, we train a sequence-to-sequence model using a span-infilling objective (\S~\ref{sec:vae_pretrain}). Second, we attach a bottleneck to this model and train it as an autoencoder for sequence compression (\S~\ref{sec:mqae}). Third, we further tighten the bottleneck and fine-tune the model so that it becomes a VAE (\S~\ref{sec:mqvae}). Below we provide details of the training procedure, beginning with the second stage.

\subsubsection{AE with Multi-Query Compression}
\label{sec:mqae}

% Hook paragraph
There have been several explorations of VAEs for symbolic music \cite{roberts2018hierarchical,yang2019deep,jiang2020transformer,wang2020pianotree}.  
However, none of these models meet the requirements of phrase-level sequence compression needed for our framework.  
Earlier works \cite{roberts2018hierarchical,yang2019deep,jiang2020transformer} focus primarily on monophonic sequences and therefore cannot handle polyphonic textures or multitrack inputs.  
PianoTreeVAE \cite{wang2020pianotree} supports polyphonic structures, but its latent representation is high-dimensional (512-dim) and, as we will show later, such a large latent size becomes problematic for diffusion training.  
Across all these VAEs, two additional limitations remain:  
(1) none achieve sufficiently accurate reconstruction for high-fidelity modeling, and  
(2) instrument identity is not explicitly encoded in the latent representation.

In contrast, our framework requires a latent representation that simultaneously satisfies five criteria:  
(1) high-fidelity reconstruction to support generation naturalness,  
(2) ability to handle variable-length note sequences to handle instruments with different complexities,  
(3) support for polyphonic and multitrack structures,  
(4) compactness for efficient diffusion training, and  
(5) explicit encoding of instrument information.

% Model structure
To ensure reconstruction quality, we start with building an autoencoder (AE) without KL regularization \cite{kingma2013auto}. Our AE is a sequence-to-sequence model implemented with an encoder–decoder Transformer \cite{vaswani2017attention}. The input and output of the model are the same \textit{phrase sequence} (REMI-z track sequence), since the model learns to compress and reconstruct it. As in standard encoder–decoder setups, the encoder produces a sequence of hidden states, and the decoder autoregressively reconstructs the sequence during inference. The difference is that the decoder does not attend to the full encoder states. Instead, it receives only a compressed representation of hidden states through a bottleneck.

% Previous method of information aggregation
Previous work in NLP has tried to obtain sequence-level information using methods such as the CLS token in BERT \cite{devlin2018bert}, sequence-level pooling \cite{reimers2019sentence}, or cross attention with single external query \cite{wang2019t} for classification tasks. These methods do capture some high-level semantics, but they are not sufficient for accurate sequence reconstruction. Compressing all information into a single latent vector is simply too difficult to retain all fine-grained details in the sequence to be compressed. 
% for our task, since full reconstruction requires complete detailed information.

% Our solution
Our solution is straightforward: instead of compressing the entire sequence into a single vector, we compress it into $m$ vectors. This gives us a way to balance completeness and compactness. A larger $m$ preserves more information but is less compact, and a smaller $m$ is more compact but loses detail. In our experiments, we set $m = 4$. This is implemented by introducing multiple special queries that attend to the encoder hidden states. Each query performs dot-product attention over the sequence and extracts a different aspect of the information. The decoder then uses these $m$ queried outputs for cross-attention.

\subsubsection{Pretraining with Span Infilling}
\label{sec:vae_pretrain}

% Intuition
There is an important component of music sequence modeling that is not directly related to compression, namely the sequence representation itself, i.e., what a valid music sequence looks like. Since we adopt the REMI-z tokenization, both the input sequence to be compressed and the output sequence to be reconstructed must follow its grammar. For example, each phrase sequence must start with an instrument token, and a duration token must always appear after the corresponding pitch token. Learning this grammar does not require compression training, and we hypothesize that learning to generate sequences that follow this grammar is helpful for later compression, because the model can build a strong contextual representation first.

% Implementation
This idea is implemented as a pretraining stage before the multi-query compression training. We use a denoising autoencoder setup similar to the BART training framework \cite{lewis2019bart}, where the objective is span infilling. During training, several spans of random length are masked out from the input sequence and fed into the encoder. The decoder is then asked to autoregressively generate the full sequence while correctly recovering the masked regions. Through this process, the model learns robust and contextual token representations and becomes able to generate grammatically correct sequences under the REMI-z rules. As noted in the original BART paper, such pretraining is very effective for sequence generation tasks, which is why we adopt it here. This pretraining stage is performed without any bottleneck, which means the decoder can see the entire hidden states sequence produced by the encoder. After the pretraining, the model is later fine-tuned with the multi-query bottleneck for compression.

\subsubsection{VAE with Progressive Bottleneck}
\label{sec:mqvae}

% Motivation: latent space need to be regularized
The autoencoder described above is not the final model, because its latent space is not well regularized. If we plot the latent codes using 2D or 3D PCA, the space does not appear symmetric, and the values of different dimensions vary widely. This is not ideal for training the latent diffusion model, which assumes that the data to be modeled follows a standard normal distribution. In other domains, especially image generation \cite{rombach2022high}, VAE-based compression is commonly used to make the latent space smoother and closer to a normal distribution. This motivates us to convert our autoencoder into a VAE.

% VAE finetune
The next step is hence to fine-tune the autoencoder into a VAE. We continue training from the autoencoder checkpoint while adding a tighter bottleneck for more aggressive compression. The implementation follows the standard VAE setup. The four query outputs (512-dim) are concatenated into one long vector (2048-dim) and projected into a 128-dimensional vector. This vector is then split into two halves: one representing the mean and the other representing the standard deviation. During training, the latent is sampled using the reparameterization trick, and the KL divergence loss is used to regularize the latent distribution toward a standard normal distribution. We adopt a loss function similar to the one used in \textit{beta}-VAE, where the total loss is a weighted sum of the reconstruction loss and the KL divergence loss. In our training, we set the weight for the KL term to 0.01. (This does not mean $\beta < 1$. See Discussion for details.) As a result, we obtain 64-dimension well regularized and noise-robust lantent thats represents phrase sequences.

\subsection{PhraseLDM}

% Overview of PhraseLDM
After obtaining PhraseVAE, we use it to compute the latent representations for the entire dataset and train PhraseLDM on these latents. During training, the latent diffusion model still respects the REMI-z grammar at a higher level, i.e. each bar contains multiple phrases ordered by instrument pitch, end with a special end-of-bar token, and then the phrases of the next bar follow. This defines the data structure of the phrase-level latent sequence. Note that this ordering is only a structural arrangement of the latent sequence; the diffusion model itself generates the entire latent sequence for the full song in one shot.

\subsubsection{Model Structure}

\begin{figure}[tb]
     \centering
     \includegraphics[width=0.9\textwidth]{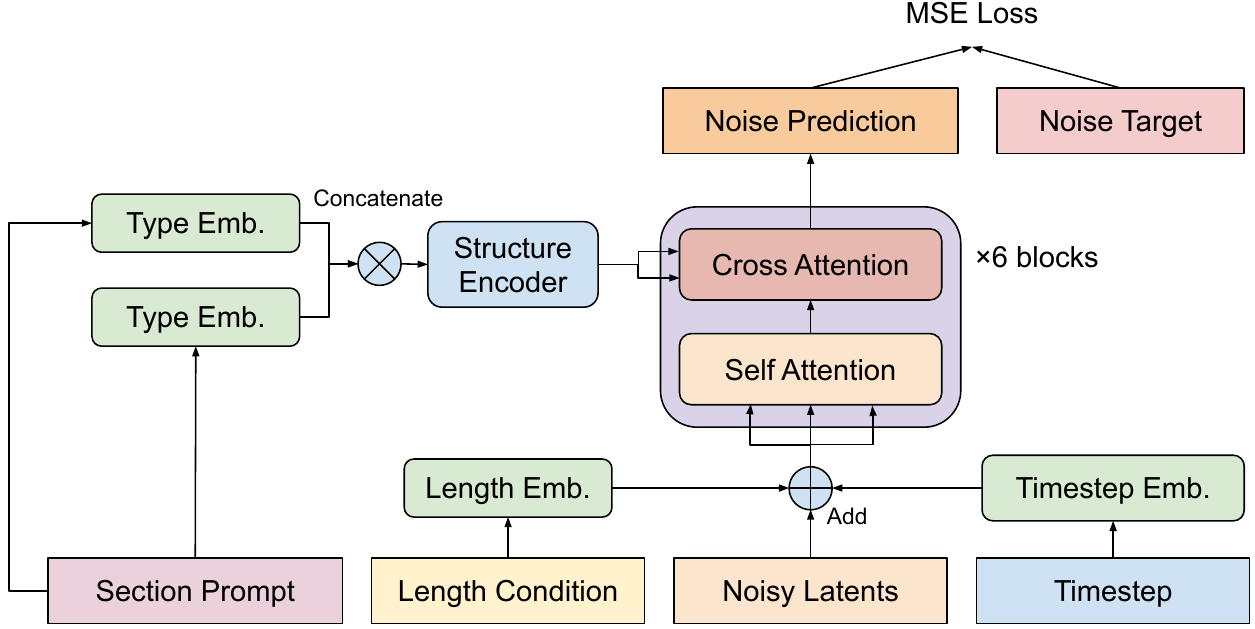}
     \caption{PhraseLDM model structure.}
     \label{fig:ldm}
\end{figure}

% Model structure
The model structure follows the design of Stable Audio \cite{evans2025stable}, which is a diffusion Transformer (DiT) \cite{peebles2023scalable} operating on the phrase latent sequence. The Transformer network has six layers and uses an encoder-only format. Each Transformer block contains both self-attention and cross-attention. Cross-attention allows the model to attend to external information and is used for conditional generation. For both training and inference, we use a standard DDPM scheduler \cite{ho2020denoising}. During training, the model predicts the noise added at each time step and the loss is computed using mean squared error (MSE). Figure~\ref{fig:ldm} shows the PhraseLDM model structure.

% Details to achieve variable length generation
There are several implementation details for handling variable-length songs. During training, all latent sequences are padded to the maximum supported length, which is 512 latents (128 bars). This allows the model to generate a fixed-length latent sequence, while different songs may have different actual lengths. To let the model learn where a song should end, we introduce a special end-of-song latent in the VAE training stage, represented by a dedicated token. When training the diffusion model, the latent corresponding to this end-of-song phrase serves as the boundary indicator. If a song is shorter than the maximum length, we pad the remaining positions using this special latent as well. During decoding, the output token sequence from the VAE is truncated at the first occurrence of the end-of-song token. We do not apply attention masks inside the LDM's self attention. Positional encoding is crucial for indicating the location and ordering of latents, and we adopt rotary position embeddings following Stable Audio.

% Conditions
The latent diffusion model receives three types of conditions during training and inference. The first is the diffusion time step, necessary for all diffusion models, which is mapped into an embedding and added to all hidden states. The other two conditions—length condition and structure condition—are optional and will be described in the following subsections.

\subsubsection{Length Condition}

% Motivation of length condition
At the initial stage of training, we found that the unconditional model sometimes struggles to maintain consistent phrases across bars. We hypothesize that this is because the distribution of phrase latents may differ across songs of different lengths. For example, longer songs may contain more repetitions or recurring patterns. Hence let the model to learn length-conditioned distribution might alleviate this issue. Adding a length condition also improves usability, since users may want to generate songs of specific lengths for different use cases.

% How length condition is implemented
We provide the length information using the number of bars in the song both for training and inference. This value is mapped to an embedding and added to all latent vectors in the sequence. To avoid overfitting to specific bar counts, we use a length bucket instead of the exact number. The condition specifies a target range of bars, where the bucket size is $B$, and each range is a multiple of $B$. In our experiments, we set $B = 10$. For example, a length condition may look like “generate a song that contains bars in the range $[40, 50)$.”

\subsubsection{Structure Condition}

% [Motivation of structure conditioning]
There are two motivations for adding structure conditioning. The first is to make the system more usable, since some users may want the generation to follow a predefined structure. The second is to enable a more clear form structure in the generation. We observe although the length-conditioned model already shows a certain degree of section structure on its own, the section boundaries and recurrence patterns are still not as clear as in real songs. Therefore, we think that providing external section information as a prompt may further improve structure-related generation quality.

% [How structure condition is implemented]
In our experiment, we use the “section layout” as the structure prompt. This layout comes from human annotations, which is a list of structure descriptor and each element describe the type and length of one section in the song, in sequential order, such as: [intro - 8 bars, verse - 10 bars, chorus - 8 bars, outro - 4 bars]”. The section type and the section length are tokenized and embedded separately, then concatenated, and sent to a small Transformer encoder to make the prompt more contextualized. The resulting embedding is then fed into the cross-attention of the DiT.

\section{Experiments}

\subsection{Dataset}

% Dataset intro
We adopt the POP909 dataset \cite{wang2020pop909} for both training and evaluation. We randomly select 5\% of the songs as the validation set and use the remaining pieces for training. POP909 provides a convenient multitrack setting with moderate musical complexity: each song contains three tracks---a melody line (monophonic), a lead instrument line (sometimes polyphonic), and a piano accompaniment (polyphonic)---which makes it well-suited for testing our proposed framework.

One limitation of POP909 is its relatively small size, containing only about 800 songs after removing songs with annotation errors. Nevertheless, we will later show that our approach is capable of achieving high-quality, diverse, and musically coherent full-song generation even under such limited data conditions, with minimal signs of overfitting.

\subsection{PhraseVAE}

\subsubsection{Implementation Details}
PhraseVAE is implemented using the encoder and decoder modules from the T5 model \cite{raffel2020exploring} as provided in the HuggingFace Transformers library \cite{wolf2019huggingface}. Both the encoder and decoder contain 3 layers with a hidden size of 512, 6 attention heads, and a positionwise feed-forward inner dimension of 1024, resulting in a model of approximately 15M parameters, comparable to our baseline VAE \cite{wang2020pianotree}. The multi-query bottleneck uses four learnable 512-dimensional query vectors whose attended outputs are concatenated and projected into a 64-dimensional VAE latent. Training uses a batch size of 128 and a learning rate of \(1\times10^{-4}\), with early stopping (patience of 20 epochs) applied at each stage and the final checkpoint selected based on the lowest validation loss.

% \subsubsection{Implementation Details}
% % Implementation details of PhraseVAE
% PhraseVAE is implemented as an encoder--decoder Transformer following the T5 implementation \cite{raffel2020exploring} provided by HuggingFace Transformers \cite{wolf2019huggingface}. Both the encoder and decoder contain 3 layers, with a hidden size of 512, 6 attention heads, and a positionwise feed-forward inner dimension of 1024. The resulting model has approximately 15M parameters, comparable in scale to our baseline \cite{wang2020pianotree}. The multi-query bottleneck is 512x4 dimension, and VAE bottleneck is 64 dimension. For training, we use a batch size of 128 and a learning rate of \(1\times10^{-4}\). Across all training stages, models are optimized until the validation loss plateaus, with early stopping (patience 20 epochs) and checkpoint selection based on the lowest validation loss.

During both AE and VAE training stages, we prepend four special tokens (\texttt{Q1}--\texttt{Q4}) to the encoder input. The corresponding encoder outputs of these tokens serve as the compressed latent vectors and are provided to the decoder as its cross-attention inputs.

\subsubsection{AE for Phrase Sequence Compression}

This part of experiments evaluate models undergone the span infilling pretrain and the AE training stages, before the VAE finetuning.

\paragraph{F1-based Metrics.}
% Introducing F1-based metrics
When evaluating reconstruction quality for both the AE and VAE, our primary concern is fidelity: the reconstructed sequence should match the original phrase as closely as possible. Many similarity metrics could be used, but we choose one that aligns naturally with musical interpretation. We adopt the F1 score as our main metric, where both missing notes and spurious notes are penalized. Concretely, the reconstructed REMI-z track sequence is converted into a bar-level piano roll with the same 48th-note quantization, and F1 scores are computed from this representation.

% Types of F1 adopted
We report three levels of F1 score with progressively stricter correctness criteria: 
\(\mathbf{F1}_{\mathrm{op}}\), \(\mathbf{F1}_{\mathrm{opd}}\), and \(\mathbf{F1}_{\mathrm{iopd}}\).
A predicted note is counted as correct if (1) onset and pitch match (\(\mathrm{F1}_{op}\)),  
(2) onset, pitch, and duration match (\(\mathrm{F1}_{opd}\)), or  
(3) instrument, onset, pitch, and duration all match (\(\mathrm{F1}_{iopd}\)).

Most prior works evaluate reconstruction by comparing individual note attributes separately (e.g., onset only or pitch), rather than assessing all attributes jointly. We argue that evaluating multiple attributes together provides a more meaningful measure of musical fidelity, and therefore adopt the three composite metrics introduced above. In practice, our model achieves \(\mathrm{F1}_{iopd}\) scores that are nearly identical or highly comparable to its \(\mathrm{F1}_{opd}\) scores, indicating that instrument identity is reliably preserved. For this reason, we primarily report \(\mathrm{F1}_{op}\) and \(\mathrm{F1}_{opd}\) in most comparisons.

% Results
\paragraph{AE Performance}
On the phrase compression task, our multi-query compression AE model achieves 
\textbf{99.9\% \(\mathrm{F1}_{op}\)} and \textbf{99.9\% \(\mathrm{F1}_{opd}\)}.

\paragraph{Comparison of Compression Strategies for AE.}

\begin{table}[t]
\caption{Comparison of different AE compression strategies (bottlenecks).}
\label{tab:ae}
\centering

\small
% \resizebox{\columnwidth}{!}{%
\begin{tabular}{@{}lccc@{}}
\toprule
\multicolumn{1}{c}{\textbf{Model name}} & \(\mathbf{F1}_{\mathrm{op}}\) & \(\mathbf{F1}_{\mathrm{opd}}\) & \textbf{Best step} \\ \midrule
Pooling-based & 98.5\% & 97.3\% & - \\
Query-based (m=1) & 93.1\% & 89.1\% & - \\
Multi-query (m=4) & \textbf{99.7\%} & \textbf{99.5\%} & \textbf{119} \\
w/o pretrain & 99.6\% & 99.1\% & 143 \\ \bottomrule
\end{tabular}%
% }

\end{table}

% Comparison experiment setting
In a more challenging setting, we further compare different compression strategies to understand their effectiveness. We consider two commonly used baselines: (1) \textit{sequence pooling}, which applies average pooling over all encoder hidden states to obtain a single sequence embedding, and (2) \textit{single-query compression}, which uses cross attention with one learned query token to attend over encoder states and aggregate information into a single vector. We also include an ablation baseline that removes the span-infilling pretraining stage and trains the autoencoder directly.  
Importantly, this comparison is conducted on the more difficult \textit{bar-compression} task, where the model must reconstruct the full REMI-z bar sequence, including multiple instrument-specific phrase sequences arranged in a specific order. This is significantly harder than phrase compression, making performance differences more observable.

% Result of comparison
As in Table~\ref{tab:ae}, single-query compression performs the worst, while pooling-based compression yields substantially better reconstruction, consistent with observations in prior work. However, pooling still loses nontrivial information, achieving only \(97.3\%~\mathrm{F1}_{opd}\), leaving room for improvement. Simply increasing the number of query vectors from one to four—our proposed multi-query compression—significantly improves reconstruction quality and surpasses the pooling baseline, reaching near-perfect performance (\(99.5\%~\mathrm{F1}_{opd}\)).  
Removing the span-infilling pretraining also degrades performance and slows convergence, confirming that learning the REMI-z grammar beforehand meaningfully benefits downstream compression.

\subsubsection{VAE with Bottleneck}

\begin{table}[tb]
\caption{Comparison of VAE bottleneck size.}
\label{tab:vae_bottleneck}
\centering
\small
\begin{tabular}{@{}rcc@{}}
\toprule
\multicolumn{1}{c}{\textbf{Latent dim}} & \(\mathbf{F1}_{\mathrm{op}}\) & \(\mathbf{F1}_{\mathrm{opd}}\) \\ \midrule
128 & 99.9\% & 99.7\% \\
\textbf{64} & \textbf{99.0\%} & \textbf{98.4\%} \\
32 & 95.1\% & 92.6\% \\ \bottomrule
\end{tabular}%

\end{table}

% ↑↓

\paragraph{Latent Size.}
Recall that in the VAE finetuning, we add a tighter bottleneck on top of the multi-query bottleneck (2048-dim per phrase sequence) to further tighten the latent space. We trained PhraseVAEs with different VAE bottleneck dimensions to produce latents of various sizes, and here we present the comparison in Table~\ref{tab:vae_bottleneck}.

Models with 128-dimensional latents or larger can almost perfectly reconstruct the input music; however, as we will show later, such large latent spaces are suboptimal because they make the subsequent latent diffusion model prone to memorization. In contrast, a 32-dimensional bottleneck is overly restrictive, achieving only 95.1\% \(\mathbf{F1}_{\mathrm{op}}\). In symbolic music, onset or pitch errors are highly perceptible: even small inaccuracies can cause notes to shift to incorrect timings or pitches, rendering the reconstruction immediately musically unacceptable. For this reason, we do not adopt a 32-dimensional latent space. Instead, we choose a 64-dimensional bottleneck, which remains compact while still providing high reconstruction fidelity.

\paragraph{Representation Granularity.}

% Introducing the motivation
This comparison shows that phrase-level compression is not only conceptually meaningful for downstream generation, as discussed earlier, but also empirically the most efficient granularity for LDM. It strikes a balance between semantic completeness and latent compactness, making it an effective “sweet spot’’ for symbolic music representation.

% Introducing baselines
We compare our PhraseVAE against two alternative granularities:

\begin{itemize}
    \item \textbf{BarVAE.} The VAE compresses the entire REMI-z bar sequence into a single latent vector. One latent encodes all instruments and all notes within the bar. This representation is highly expressive because it preserves complete bar-level musical information.
    \item \textbf{PositionVAE.} The VAE compresses each active temporal position inside a bar. For every non-empty 48th-note slot that contains one or more note onsets (across all instruments), the model produces one latent. This representation is more fine-grained than PhraseVAE. 
\end{itemize}

% Experiment setting
As shown earlier, more compact latents make accurate reconstruction harder. Conversely, by increasing the latent dimensionality sufficiently, one can always obtain a VAE that reconstructs nearly perfectly. However, this is not desirable: large latent vectors directly increase the diffusion model’s channel size, and prior work in latent diffusion suggests that high-dimensional latents make LDM training more difficult. Therefore, for each representation granularity, we search for the \textit{smallest} bottleneck size within \{16, 32, 64, 128, 256, 512, 1024\} that achieves \(\mathbf{F1}_{\mathrm{opd}} > 98\%\), and report the resulting information footprint in Table~\ref{tab:vae_granularity}.

\begin{table}[tb]
\caption{Comparison of latent granularity.}
\label{tab:vae_granularity}
\centering
\small
\begin{tabular}{@{}lrrrc@{}}
\toprule
\textbf{Model} & \multicolumn{1}{l}{\textbf{Latent size per unit}} & \multicolumn{1}{l}{\textbf{Avg. \#units per bar}} & \multicolumn{1}{l}{\textbf{Avg. latent dim per bar}} & \multicolumn{1}{l}{\(\mathbf{F1}_{\mathrm{opd}}\)} \\ \midrule
BarVAE & 512 & 1 & 512.0 & 98.3\% \\
PositionVAE & 32 & 9.54 & 305.3 & 98.7\% \\
PhraseVAE & 64 & 2.38 & \textbf{152.3} & 98.4\% \\ \bottomrule
\end{tabular}%

\end{table}

% Result analysis
Although the three models achieve similar reconstruction quality, they differ substantially in efficiency for LDM training. This is because the LDM must generate a different number of latent vectors per bar depending on the granularity:

\begin{itemize}
    \item \textbf{BarVAE:} one latent per bar, resulting in \(512\) floating-point values.
    \item \textbf{PositionVAE:} roughly \(10\) non-empty positions per bar, leading to around \(300\) floating-point values.
    \item \textbf{PhraseVAE:} on average \(2.38\) phrases per bar, resulting in about \(150\) floating-point values.
\end{itemize}

PhraseVAE requires the fewest floating-point values to represent the same bar of music, making it the most efficient design for latent diffusion among the three granularities, which potentially facilitate LDM learning. For this reason, we adopt phrase-level latents for our LDM.

\paragraph{Comparison with Existing Models.}

% Experiment setting
We evaluate the reconstruction capability of our model by comparing it against a widely adopted symbolic music VAE, PianoTreeVAE \cite{wang2020pianotree}. This comparison is performed on the bar-level compression task, which is more challenging than phrase-level compression. We retrained PianoTreeVAE on our bar-level dataset after flattening all instruments (i.e., treating all notes as belonging to a single instrument), and adopted the same temporal quantization as our system, namely 48th-note resolution for onset and duration. For a fair comparison, we used a 512-dimensional latent vector, matching the latent size of PianoTreeVAE.

\begin{table}[tb]
\caption{Comparision between our VAE and PianoTreeVAE.}
\label{tab:vae_vs_pianotree}

\centering
\small
\begin{tabular}{@{}lccc@{}}
\toprule
\multicolumn{1}{c}{\textbf{Model}} & \(\mathbf{F1}_{\mathrm{op}}\) & \(\mathbf{F1}_{\mathrm{opd}}\) & \textbf{Training Time} \\ \midrule
PianoTreeVAE & 92.2\% & 88.7\% & 30 hours \\
Ours (512-dim BarVAE) & 98.9\% & 98.3\% & 14 hours \\ \bottomrule
\end{tabular}%

\end{table}

% Baseline is bad and ours is good
As in Table~\ref{tab:vae_vs_pianotree}, our model substantially outperforms PianoTreeVAE, achieving roughly a 10\% improvement in \(\mathbf{F1}_{\mathrm{opd}}\). Additionally, due to the parallel training capability of the Transformer architecture (as opposed to the autoregressive nature of PianoTreeVAE), the total training time of our method—including all three stages—is much shorter than that of PianoTreeVAE.

% PianoTreeVAE performs worse than the scores reported in the original paper. A key contributing factor appears to be the temporal quantization: the original implementation uses 16th-note resolution, which is too coarse to represent contemporary symbolic music that frequently contains rhythms finer than a 16th note. Modern symbolic generation research commonly adopts 48th-note resolution, and we follow this standard here.

\subsection{PhraseLDM}

\subsubsection{Implementation Details}

% Implementation details.
PhraseLDM is implemented as an encoder-only Transformer and adopts the implementation from Stable Audio DiT. The model uses a hidden dimension of 512 (\(d_{\mathrm{model}} = 512\)), 64 input and output channels, 6 Transformer layers, and 16 attention heads, resulting in a model of approximately 30M parameters. The context length is set to 512 latents, supporting up to 128 bars of music for the adopted dataset (each bar contains three phrase latents plus an end-of-bar latent). The cross-attention module operates at 128 dimensions, matching the output size of the structure encoder. Diffusion timesteps are encoded via a 128-dimensional time projection. Length conditioning is provided through a simple embedding layer that outputs 128-dimensional vectors. The structure encoder is implemented as a 3-layer Transformer encoder. For training, we use a batch size of 128 and a learning rate of \(5\times10^{-4}\). All models are trained for 200k steps.

\subsubsection{Metrics for Generation Quality}

% Overview of metrics
Since full-song symbolic music generation is a relatively new task, there is no established standard for evaluation. We therefore propose a set of metrics that we believe are reasonable and informative.

% Introducing PhraseFID
\textbf{PhraseFID.} This metric evaluates phrase-level generation quality. We directly compute FID score \cite{heusel2017gans} in the PhraseVAE latent space between latents of generated songs and that of the training set. It measures (1) how well the model generates phrases that resemble those in the training data, and (2) the phrase-level musical quality, with the assumption that phrases closer to human created phrases have higher quality. A lower PhraseFID indicates that generated phrases follow a distribution closer to real data and are hence better in quality.

\subsubsection{Metrics for Section Structures}

% Introducing BarSSM
We adopt a bar-level latent Self-Similarity Matrix (\textbf{BarSSM}) to evaluate structural quality. Existing structure-related metrics (e.g., those in \cite{wang2024whole}) apply only to structure-conditioned generation. In contrast, BarSSM measures structure coherence for any model.  
The method is inspired by the long-standing use of SSMs in the music structural segmentation task. Since repetition is a key characteristic of contemporary music and a direct indicator of section recurrence, SSMs of real songs typically display clear patterns.  
To compute BarSSM, we first average-pool the phrase latents within each bar to obtain a single ``bar latent.'' We then compute a self-similarity matrix using pairwise cosine distances between all bar latents, and visualize it as a two-dimensional image.

\begin{figure}[tb]
    \centering
    \begin{subfigure}{0.32\textwidth}
        \centering
        \includegraphics[width=\textwidth]{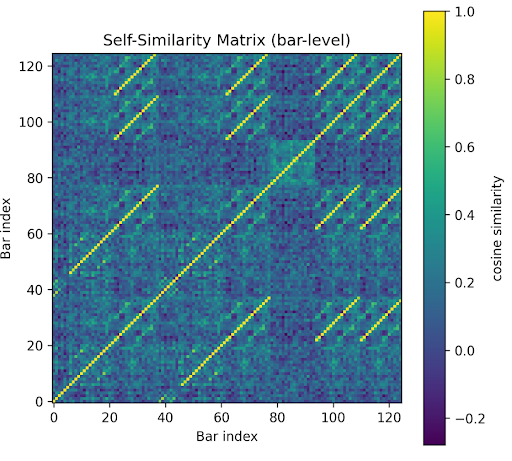}
        \caption{Real song SSM}
        \label{fig:ssm_real}
    \end{subfigure}
    \hfill
    \begin{subfigure}{0.32\textwidth}
        \centering
        \includegraphics[width=\textwidth]{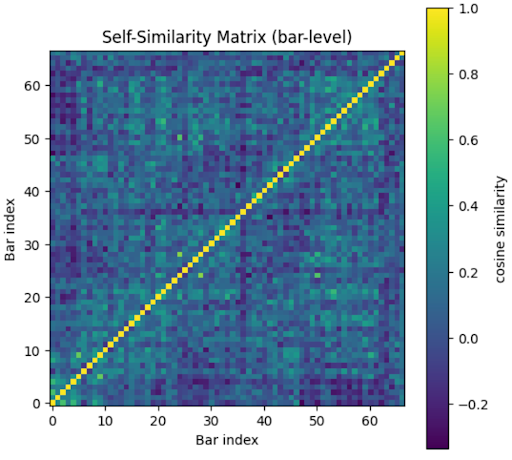}
        \caption{Weak structure in generated song}
        \label{fig:ssm_bad}
    \end{subfigure}
    \hfill
    \begin{subfigure}{0.32\textwidth}
        \centering
        \includegraphics[width=\textwidth]{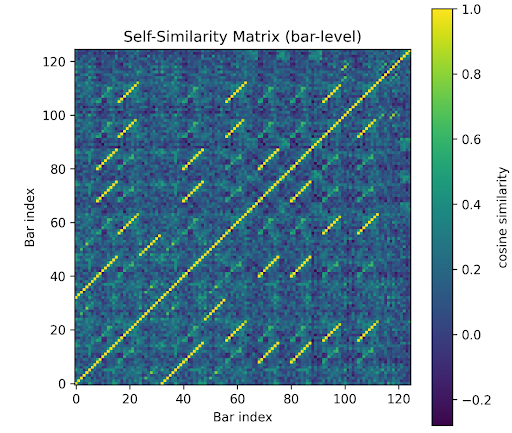}
        \caption{Clear structure in generated song}
        \label{fig:ssm_good}
    \end{subfigure}

    \caption{Examples of bar-level Self-Similarity Matrices (SSMs) from a real song, a weakly structured generated song, and a structurally coherent generated song.}
    \label{fig:ssm_example}
\end{figure}

% Explain why SSM is useful
SSMs on symbolic music latents reveal structure almost directly. As illustrated in Figure~\ref{fig:ssm_real}, genuine songs show distinctive non-major-diagonal lines formed by consecutive bright points, indicating the presence of similar sections that recur in the composition. From such a pattern, one can infer the section layout of a real song—e.g.,  
\textit{intro – A + B – interlude – A + B – bridge – B × 2}—  
without even listening to the audio.

% How did we use SSM
Thus, we visually inspect SSMs of generated samples to determine whether clear section-level concepts appear. If the SSM lacks meaningful non-major-diagonal structure, the song is likely a stream of unrelated bars with no recurrence (as in Figure~\ref{fig:ssm_bad}). In contrast, a well-trained model produces SSMs that resemble those of real songs (Figure~\ref{fig:ssm_good}).

% An objective metric from SSM
However, manual inspection is slow. Therefore, we propose a simple but intuitive metric to quantify section recurrence:

\textbf{SRS (Section Recurrence Score).}  
We count the total lengths of all non-major-diagonal lines in the \emph{upper triangular} region of the SSM, subject to two rules:  
(1) a diagonal must have length $\ge 4$ bars,  
(2) clustered diagonals (diagonals adjacent to each other) are excluded (these correspond to similar consecutive bars rather than recurring sections).  
After summing the lengths, we divide by the total number of bars in the song.  
SRS has a clear musical interpretation: on average, how many times a musical idea reappears in sections of length $\ge 4$ bars, beyond its original occurrence.  

% How to use SRS
The SRS depends on the threshold used to define ``bright points'' in the SSM. With a threshold of 0.5, real POP909 songs achieve an average SRS of around 0.6 (but vary across songs). A model that generates high-quality local texture but has no notion of section structure will score near 0.  
Higher SRS values indicate more repetition, which is generally desirable because most models we trained struggle with structural recurrence. 

% Conversely, models with excessively high SRS (even higher than real data) often show degraded PhraseFID, indicating a trade-off between repetition and phrase-level diversity.

\subsubsection{Metrics for Memorization}

% Motivation
While generated samples that are close to the training set often appear high-quality, it is equally important to ensure sufficient diversity. If a model simply replicates training examples, quality metrics may still look perfect, but the model effectively degenerates into a lookup table. Since the purpose of a generative model is to produce novel outputs, we must explicitly measure and control memorization.

% Melody based similarity evaluation
We follow the idea of melody-based similarity evaluation from \cite{wang2024whole}. A higher melody similarity between a generated song and the training set indicates stronger memorization. But unlike \cite{wang2024whole}, which computes segment-level similarity, we evaluate similarity at the full-song level and therefore introduce new metrics.

We extract the melody track from each song using REMI-z grammar and obtain two melody strings, \(\mathrm{mel\_str}_A\) and \(\mathrm{mel\_str}_B\). Melody similarity is defined as:
\[
\mathrm{Sim}_{\mathrm{mel}} = 1 - \mathrm{WER}(\mathrm{mel\_str}_A, \mathrm{mel\_str}_B).
\]

\textbf{MMR (Melody Memorization Rate).}  
For each generated song, we compute its melody similarity with all songs in the training set and take the maximum similarity, which indicates how similar generated melodies are to their closest training melodies.

\textbf{T2R (Top-2 Distance Ratio).}  
To determine whether a generated sample is ``memorized,'' we examine the distances between the generated melody and its top two closest training melodies. Let \(\mathrm{SIM}_1\) and \(\mathrm{SIM}_2\) denote the most similar and second-most similar training melodies, and let their WER distances be \(d_1\) and \(d_2\). We compute:
\[
\mathrm{T2R} = \frac{d_1}{d_2}.
\]
Following \cite{gu2023memorization}, we classify a sample as memorized if \(\mathrm{T2R} < \frac{1}{3}\). The intuition is that if a generated song is extremely close to one specific training sample yet noticeably different from all others, the model has likely reproduced that example.

\textbf{MR (Memorization Ratio).}  
For a batch of generated samples, we compute the proportion of songs classified as memorized under the above rule.

\subsubsection{Main Observations}

\begin{table}[tb]
\caption{PhraseLDM objective metrics. Len. Acc. refers to length accuracy.}
\label{tab:ldm}

\centering
\small
\begin{tabular}{@{}lrrrrr@{}}
\toprule
\textbf{Model} & \multicolumn{1}{l}{\textbf{PhraseFID}} & \multicolumn{1}{l}{\textbf{SRS}} & \multicolumn{1}{l}{\textbf{T2R}} & \multicolumn{1}{l}{\textbf{MR}} & \multicolumn{1}{l}{\textbf{Len. Acc.}} \\ \midrule
Real song & 4.24 & 0.444 & 0.937 & 5\% & - \\
Unconditioned & 3.89 & 0.161 & 0.984 & 0 & 15\% \\
Length conditioned & 3.81 & 0.165 & 0.989 & 0 & 95\% \\
Length and section conditioned & (7.65) & (0.024) & 0.983 & 0 & 100\% \\
Length conditioned (128-dim, 170k step) & 2.32 & 0.273 & 0.466 & 55\% & - \\
Length conditioned (128-dim, 140k step) & 4.44 & 0.075 & 0.872 & 20\% & - \\ \bottomrule
\end{tabular}%

\end{table}

We randomly select 100 songs from the training set as the evaluation set for PhraseFID calculation. Each model generates 20 songs, and we compute all metrics between these two groups. For the ``real song'' baseline, we randomly select another 20 songs from the training set that are not part of the evaluation set.

\textbf{LDM successfully learns the phrase-level distribution.}  
As in Table~\ref{tab:ldm}, both the unconditioned and length-conditioned versions of our 64-dimensional model achieve lower PhraseFID scores than the ``real song'' baseline. This indicates that the model captures the phrase-level latent distribution effectively.

Since all structure-conditioned samples are generated using the same structure prompt, their PhraseFID and SRS naturally differ from those of real data. Therefore, we do not compare their scores directly against other models.

\textbf{Models without structure conditioning still learn some degree of structure planning.}  
Both the unconditional model and the length-only model achieve an SRS of around 16\%, indicating that they have learned to reuse musical material across sections. From the generated examples on our demo site, one can observe recurring themes appearing in different parts of the song. But they still have some distance to the real song's SRS level.

\textbf{64-dim LDM generation exhibits high diversity.}  
Up to 200k training steps, none of the generated samples are classified as memorized, yielding a 0\% memorization rate. Combined with consistently high Top-2 Ratios, this shows that generated samples are distinct from all training songs and do not overly resemble any specific one. Interestingly, this diversity is even higher than that of real data under our metric, where one real song is flagged as ``memorized.''

\textbf{128-dim latents introduce notable memorization issues.}  
We also trained an LDM using a 128-dimensional PhraseVAE to compare with the 64-dim model. While the larger latent dimension reduces PhraseFID and increases SRS, this comes at the expense of memorization. By 140k steps, 20\% of generated samples are nearly identical to a specific training song. The memorization rate continues to rise during training and reaches 50\% by 170k steps.

\textbf{Length conditioning is effective, and further strengthened by structure conditioning.}  
Length accuracy is computed as follows: for length-conditioned models, we set the target to 64 bars (a near-median length in POP909) and consider outputs correct if their length falls within the soft constraint interval \([60,70)\).  
Without length conditioning, only 15\% of generated songs happen to fall within this range. With length conditioning, accuracy increases dramatically to 95\%. Adding structure conditioning further improves this to 100\%.

\subsubsection{Qualitative Results}

Due to time limitations, we did not conduct a systematic human evaluation. Instead, we report several qualitative observations based on representative, non–cherry-picked generated samples.

\begin{figure}[tb]
     \centering
     \begin{subfigure}{.9\textwidth}
         \centering
         \includegraphics[width=\textwidth]{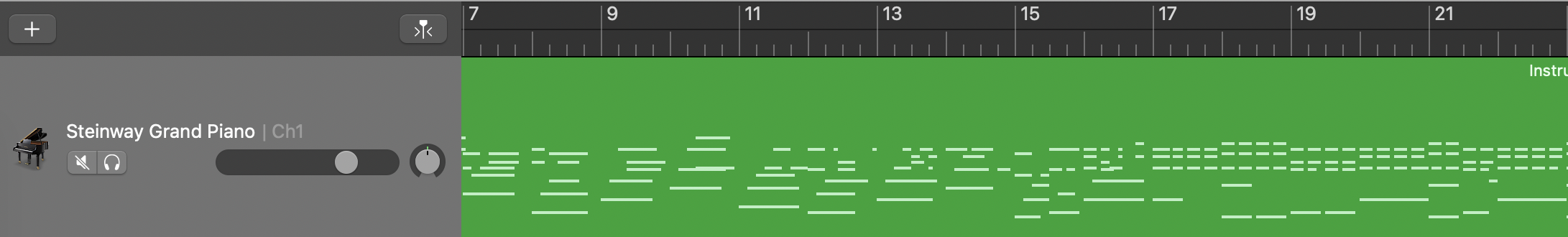}
         \caption{Transition from arpeggiated accompaniment to motoric pattern across sections. }
         \label{fig:texture1}
     \end{subfigure}
     
     \begin{subfigure}{.9\textwidth}
         \centering
         \includegraphics[width=\textwidth]{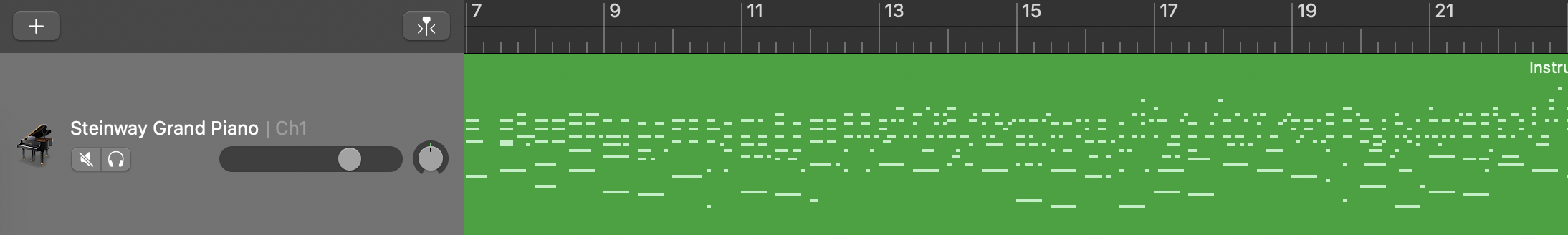}
         \caption{A texture with sophisticated rhythmic detail.}
         \label{fig:texture2}
     \end{subfigure}

     \begin{subfigure}{.9\textwidth}
         \centering
         \includegraphics[width=\textwidth]{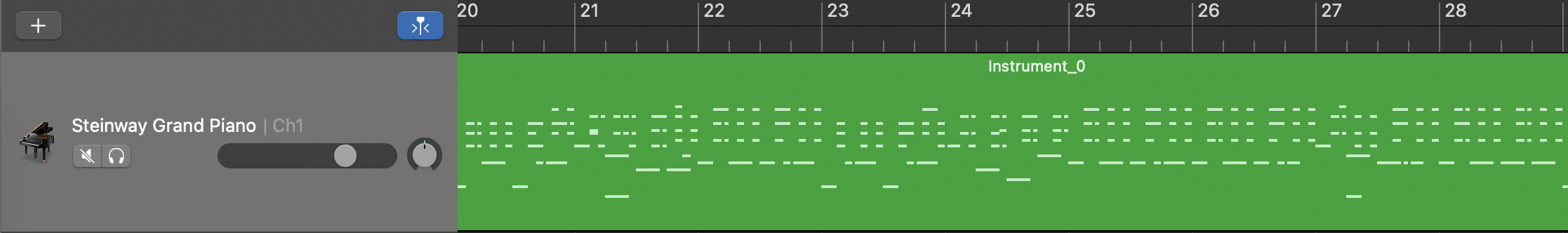}
         \caption{A bossa nova accompaniment pattern.}
         \label{fig:texture3}
     \end{subfigure}
     
     \caption{Case studies of generated piano textures. The examples demonstrate idiomatic phrasing, structural coherence, stylistic diversity, and practical playability.}
     \label{fig:piano_textures}
\end{figure}

\begin{figure}[tb]
    \centering

    \begin{subfigure}{.48\textwidth}
        \centering
        \includegraphics[width=\textwidth]{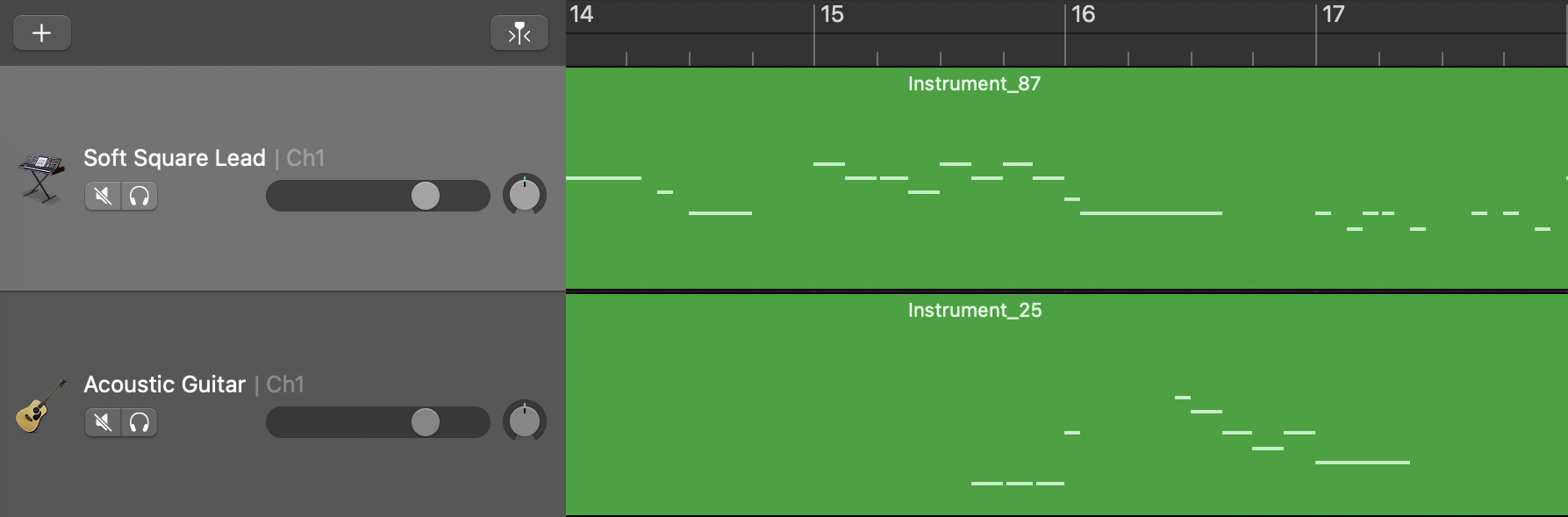}
        \caption{Lead instrument filling the spaces between melody phrases.}
        \label{fig:interact_1}
    \end{subfigure}
    \hfill
    \begin{subfigure}{.48\textwidth}
        \centering
        \includegraphics[width=\textwidth]{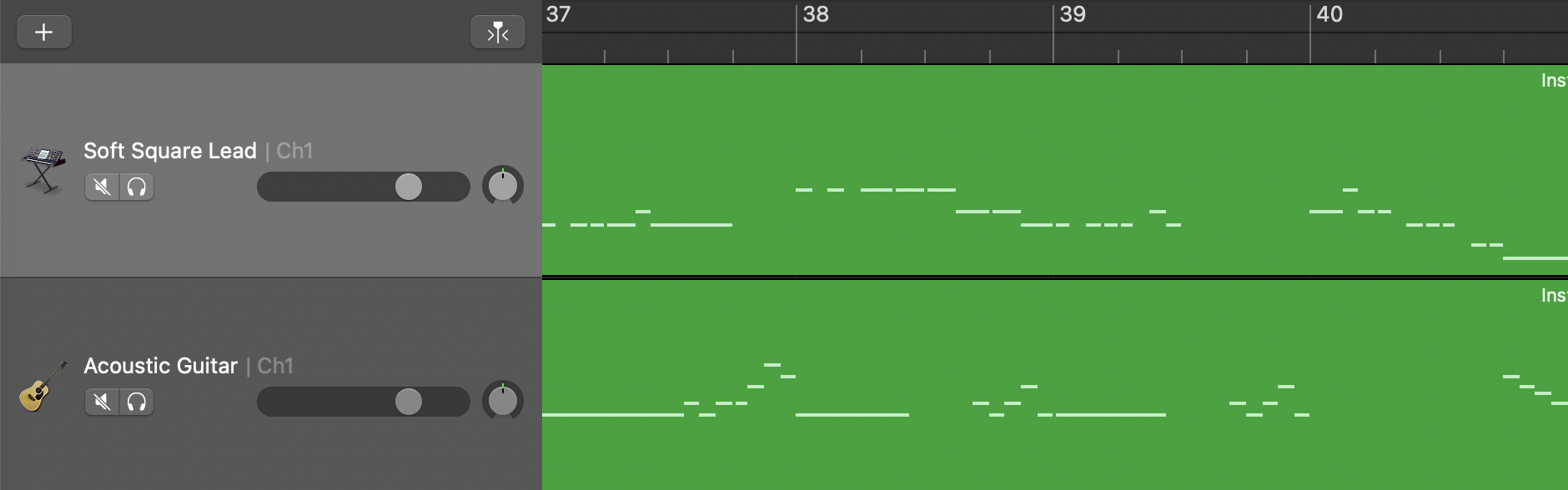}
        \caption{Lead instrument playing a countermelody that forms a rich composite texture with the melody.}
        \label{fig:interact_2}
    \end{subfigure}

    \caption{Examples of interactions between the lead instrument (the guitar track, 2nd row) and the melody track (synth lead, 1st row) in generated music.}
    \label{fig:interplay}
\end{figure}

\begin{figure}[tb]
    \centering

    % --------------------------- Intro row ---------------------------
    \begin{subfigure}{.48\textwidth}
        \centering
        \includegraphics[width=\textwidth]{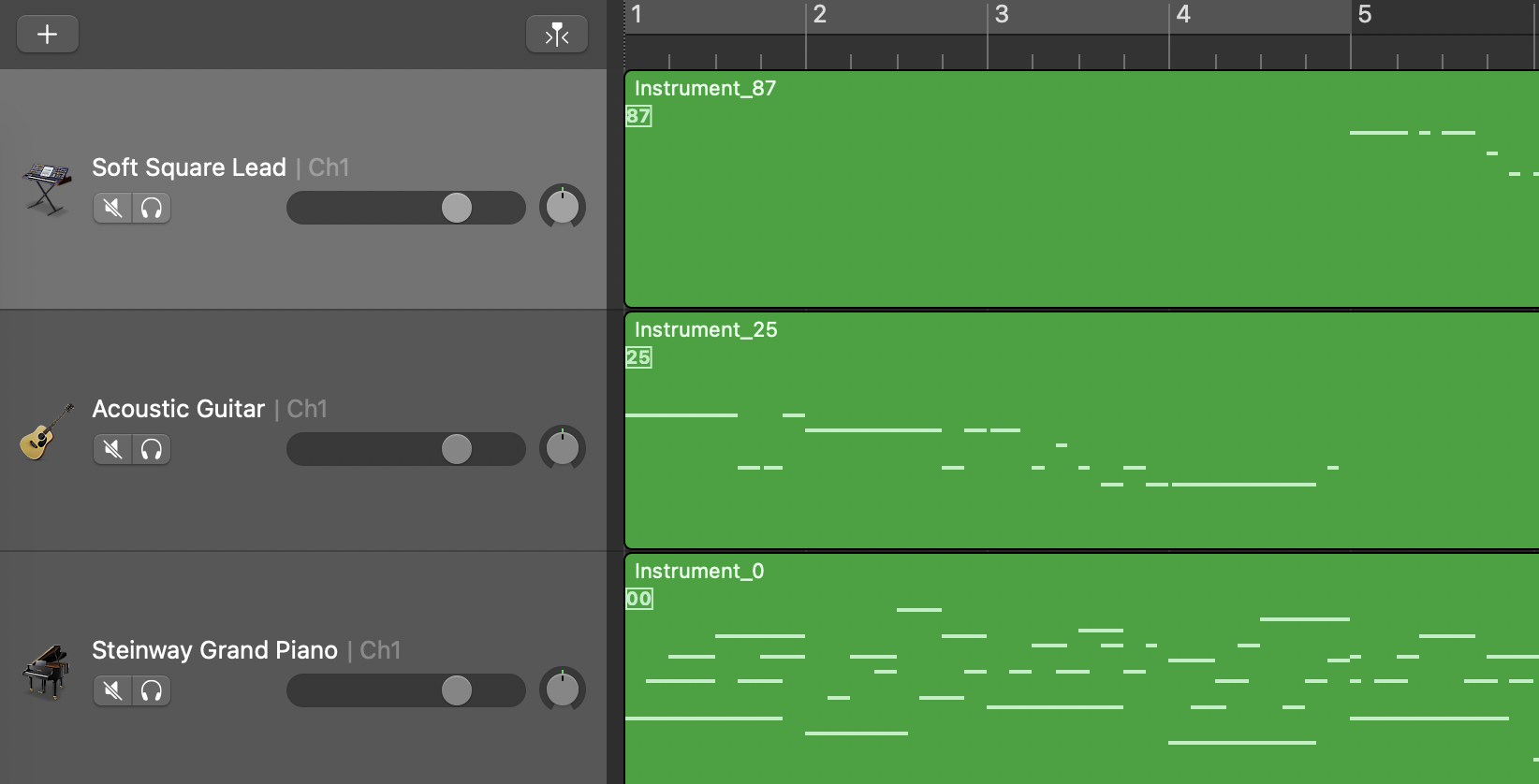}
        \caption{Intro segment with sparse texture and no melody content, reflecting a typical lead-in arrangement.}
        \label{fig:intro1}
    \end{subfigure}
    \hfill
    \begin{subfigure}{.48\textwidth}
        \centering
        \includegraphics[width=\textwidth]{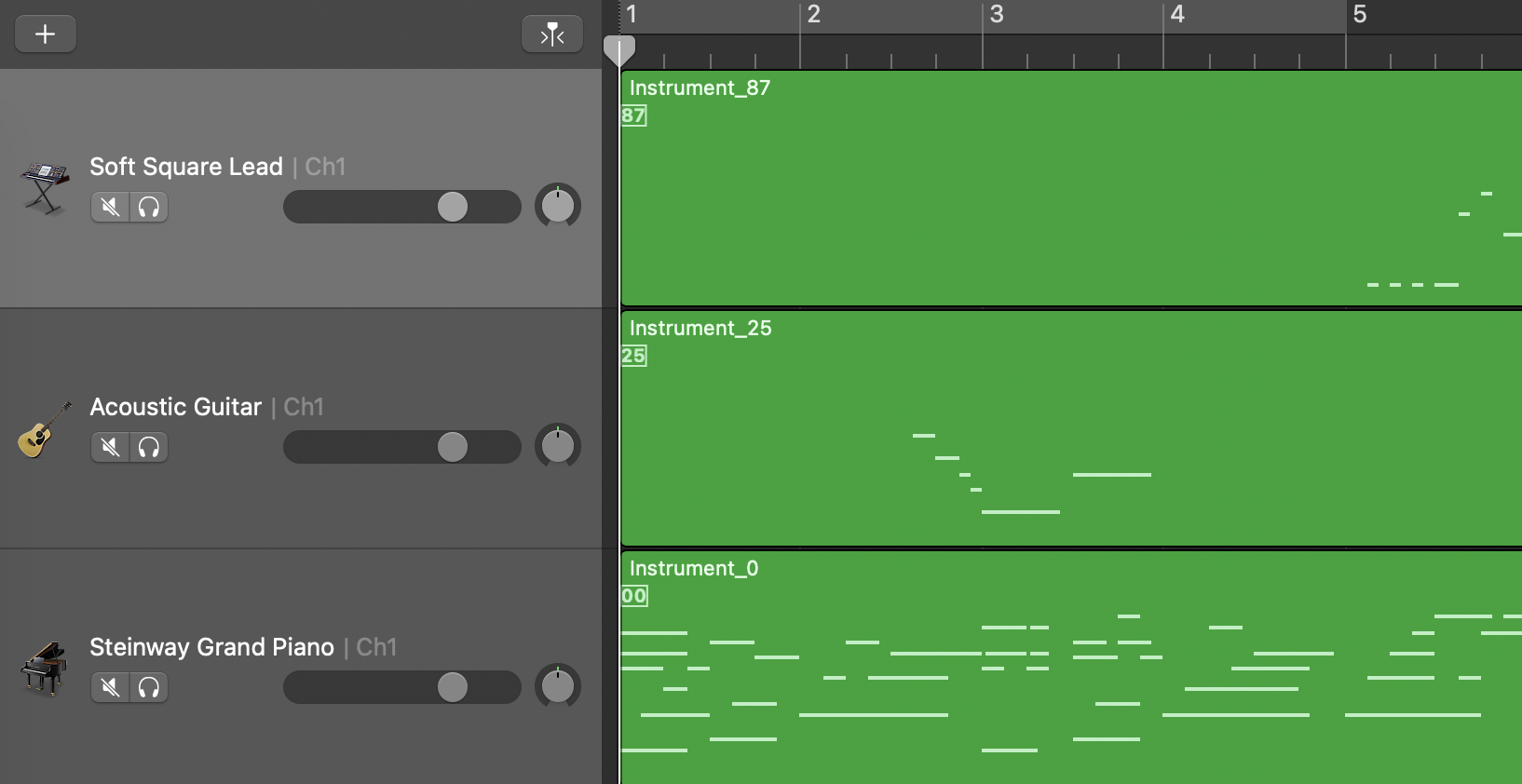}
        \caption{Another example of an intro region where the model avoids melody content in the intro part.}
        \label{fig:intro2}
    \end{subfigure}

    \vspace{1em}

    % --------------------------- Outro 1 ---------------------------
    \begin{subfigure}{.9\textwidth}
        \centering
        \includegraphics[width=\textwidth]{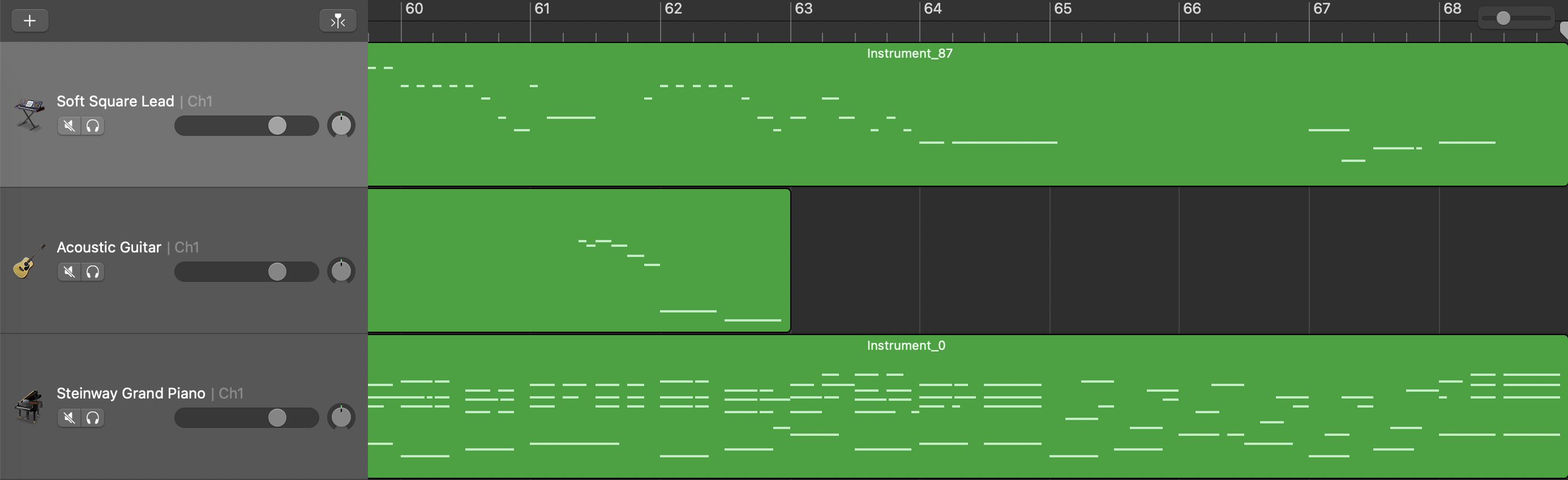}
        \caption{Outro segment demonstrating reduced texture density and simplified accompaniment, creating a natural sense of closure.}
        \label{fig:outro1}
    \end{subfigure}

    \vspace{1em}

    % --------------------------- Outro 2 ---------------------------
    \begin{subfigure}{.9\textwidth}
        \centering
        \includegraphics[width=\textwidth]{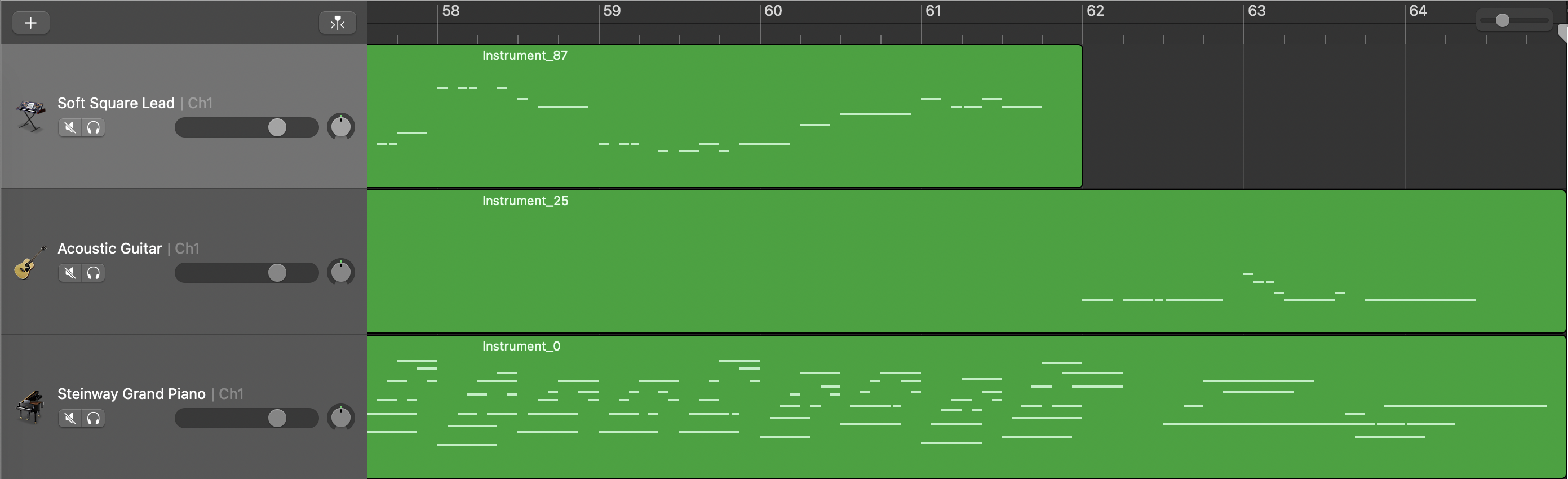}
        \caption{Another outro example where the rhythmic activity diminishes toward the end, reinforcing the ending gesture.}
        \label{fig:outro2}
    \end{subfigure}

    \caption{Examples of intro and outro regions generated by \textsc{PhraseLDM}. Intro segments consistently avoid premature melodic content, while outro segments naturally taper in texture and density, producing a convincing ending.}
    \label{fig:intro_outro_examples}
\end{figure}

\textbf{High Local Quality.}  
Segments generated by the length-conditioned model exhibit strong phrase-level musicality. The piano accompaniment patterns are idiomatic and appear perfectly playable, as in Figure~\ref{fig:piano_textures}. Chord progressions follow common POP909 conventions. Melodies sound natural and blend well with the accompaniment, and the lead instrument track often demonstrates musically meaningful behavior, such as countermelodies or interplays with the main melody, as in Figure~\ref{fig:interplay}.

% [figs: idiomatic pattern, chord progression, countermelody]

\textbf{Unconditional Model Learns Variable-Length Generation.}  
Surprisingly, the unconditional model is capable of producing full songs of widely varying lengths. For example, three randomly selected samples contain 51, 89, and 103 bars. This suggests that the model has learned the distribution of song-level latent lengths without explicit supervision.

% [figs: 51-bar, 89-bar, 103-bar samples]

\textbf{Structure-Unconditioned Models Often Produce Coherent Song-Level Structure.}  
Even without structure conditioning, the model frequently generates musically meaningful song-level structure. For instance, many samples contain clear intros and outros with characteristic sparse textures or cadential gestures, as in Figure~\ref{fig:intro_outro_examples}. Between melodic sections, the model often inserts lead-instrument fills that function as transitions. Although chorus/verse boundaries are not always as distinct as in real songs, we still observe repeated sections that indicate implicit structure planning.

% [figs: intro examples, outro examples, instrumental fills]

\begin{figure}[tb]
    \centering

    \begin{subfigure}{.95\textwidth}
        \centering
        \includegraphics[width=\textwidth]{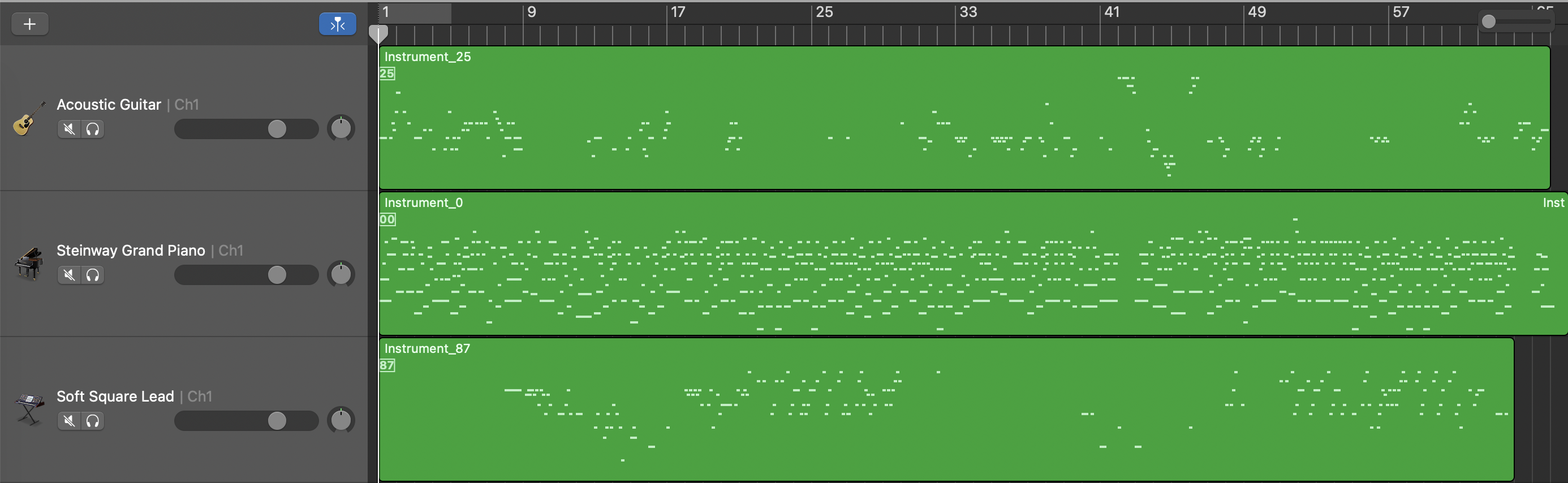}
        \label{fig:section1}
    \end{subfigure}

    \vspace{1em}

    \begin{subfigure}{.95\textwidth}
        \centering
        \includegraphics[width=\textwidth]{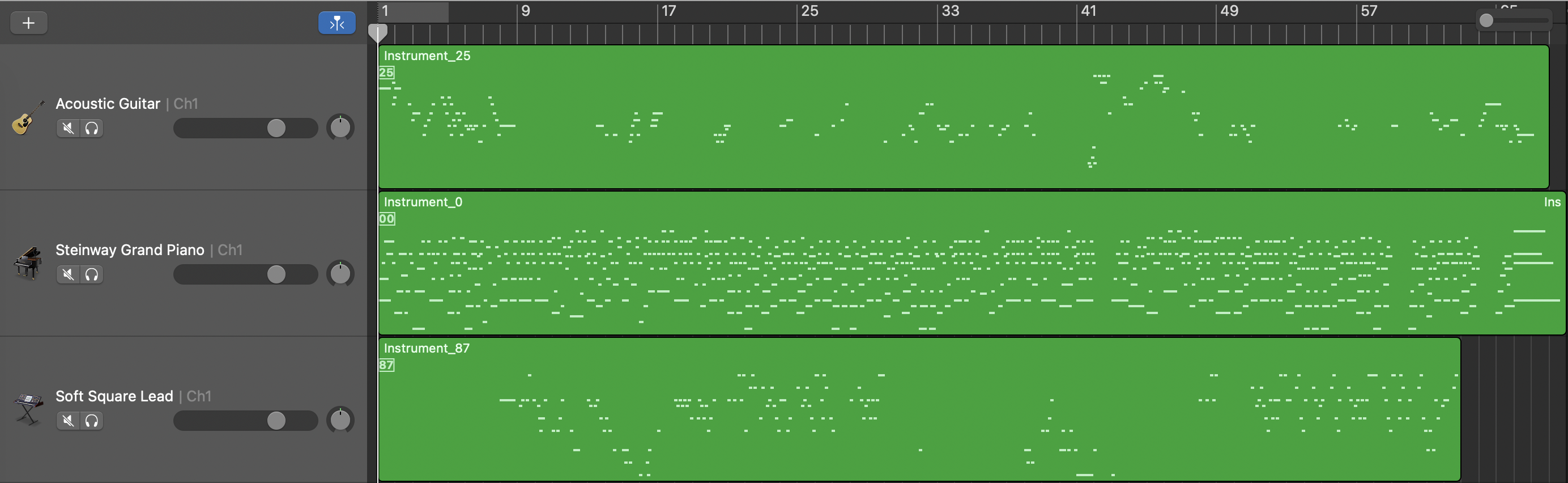}
        \label{fig:section2}
    \end{subfigure}

    \vspace{1em}

    \begin{subfigure}{.95\textwidth}
        \centering
        \includegraphics[width=\textwidth]{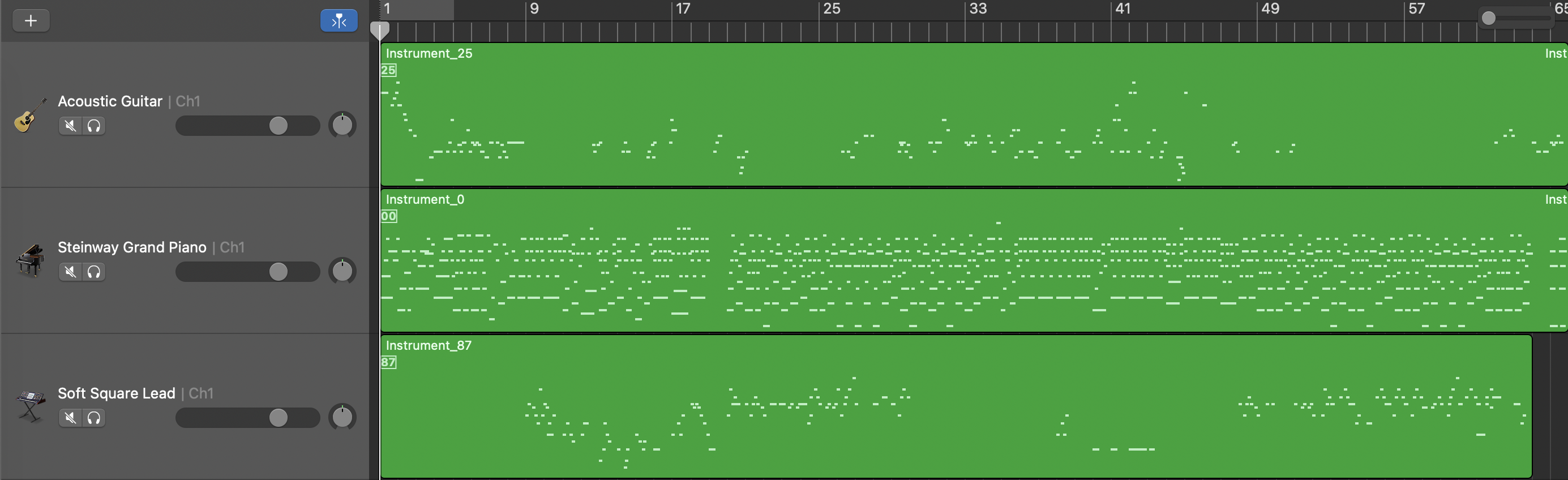}
        \label{fig:section3}
    \end{subfigure}

    \caption{Three randomly selected generations from the section-conditioned model under the same length and section conditions. The length condition is [60,70) bars. The model is conditioned on the following structure sequence:
\([\,\text{i-8},\,\text{A-8},\,\text{A-8},\,\text{B-4},\,\text{B-4},\,\text{x-4},\,\text{A-8},\,\text{B-4},\,\text{B-4},\,\text{B-4},\,\text{B-4},\,\text{X-4}\,]\).}
    \label{fig:section_conditioned}
\end{figure}

\textbf{Section conditioning is partially effective.}
When conditioned on the same length and section specification, the model consistently produces songs with similar structural layouts. As shown in Figure~\ref{fig:section_conditioned}, in the three randomly selected examples, the distributions of instrument activity exhibit clear alignment among different generated songs, indicating that the conditioning signal is indeed utilized. However, notable limitations remain. In particular, when the condition requires the second occurrence of the verse section (A-8 after x-4), all outputs fail to generate coherent melodic content. This suggests that while the model learns coarse structural placement, it does not fully capture the functional meaning of each section token. Further refinement of the section representation and conditioning mechanism is left for future work.

% [figs: structure-conditioned outputs]

Overall, qualitative inspection aligns well with our quantitative metrics and indicates that PhraseLDM produces musically coherent, diverse, and structurally meaningful full-song generations.
\section{Discussion}

\subsection{Beta Value for PhraseVAE}
In PhraseVAE, our use of a KL weight of \(0.01\) should not be interpreted as weak regularization. Because reconstruction loss is computed at the \emph{token} level while KL divergence is applied once per \emph{sequence}, setting the KL weight to \(1\) would yield an excessively large effective \(\beta\)-value and prevent the VAE from converging. This mismatch arises because a single KL term must implicitly balance against hundreds of token-level reconstruction terms; in other words, the sequence-level KL must effectively be “distributed” across all tokens to match their scale.

Despite the seemingly small weight, the resulting latent space is in fact strongly regularized: the empirical standard deviation is approximately \(0.76\), which is characteristic of a \(\beta\)-VAE with an effective \(\beta > 1\). The exact effective \(\beta\) is not explicitly defined, as the compressed sequences vary in length and therefore in the number of reconstruction terms.

\subsection{Latent Space Smoothness and Interpolation Behavior}

\begin{figure}[tb]
    \centering

    \begin{minipage}{0.45\textwidth}
        \centering
        \begin{subfigure}{\textwidth}
            \includegraphics[width=\textwidth]{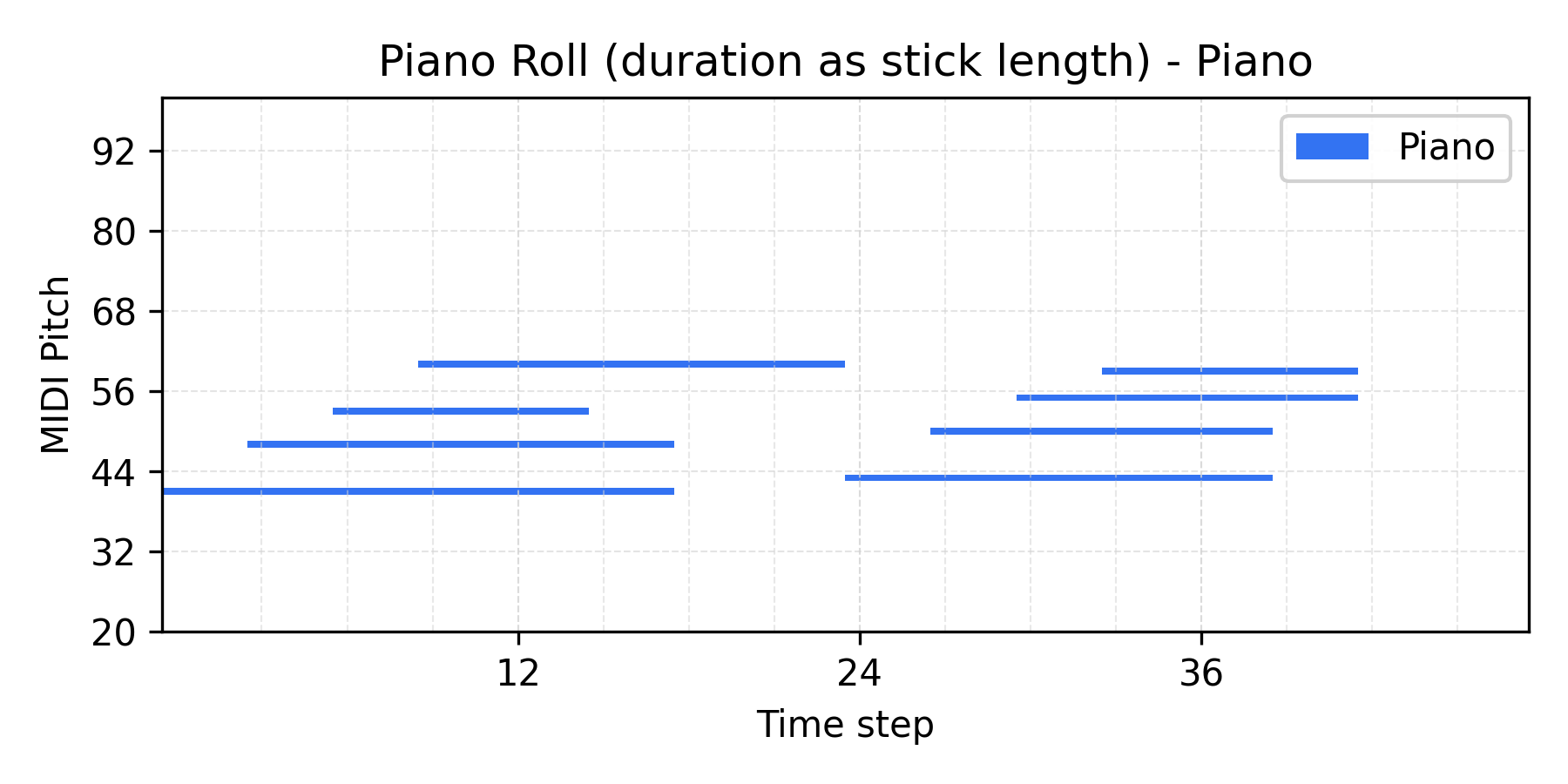}
            \caption{$\alpha = 0.0$}
        \end{subfigure}

        \vspace{0.5em}

        \begin{subfigure}{\textwidth}
            \includegraphics[width=\textwidth]{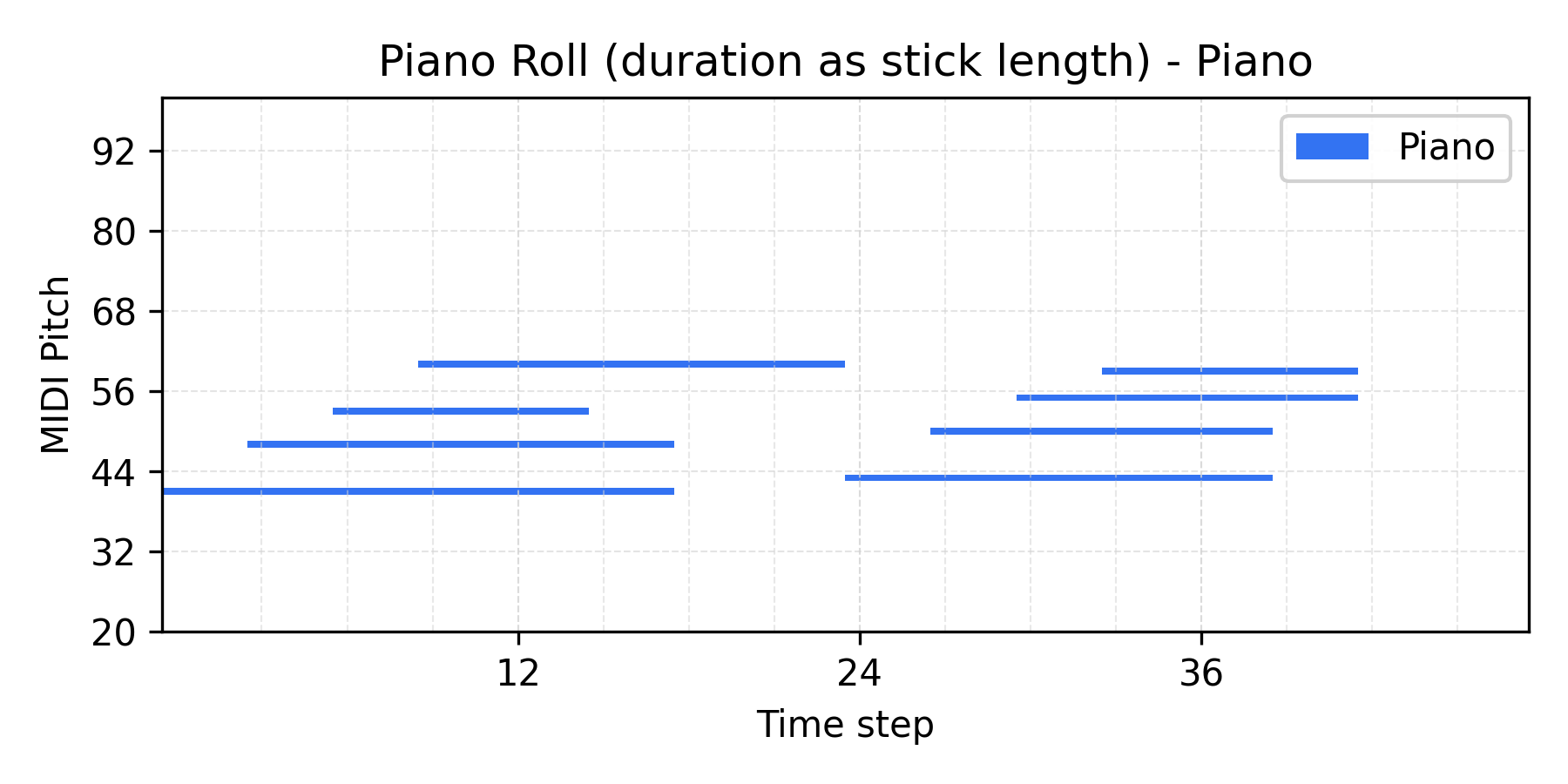}
            \caption{$\alpha = 0.3$}
        \end{subfigure}

        \vspace{0.5em}

        \begin{subfigure}{\textwidth}
            \includegraphics[width=\textwidth]{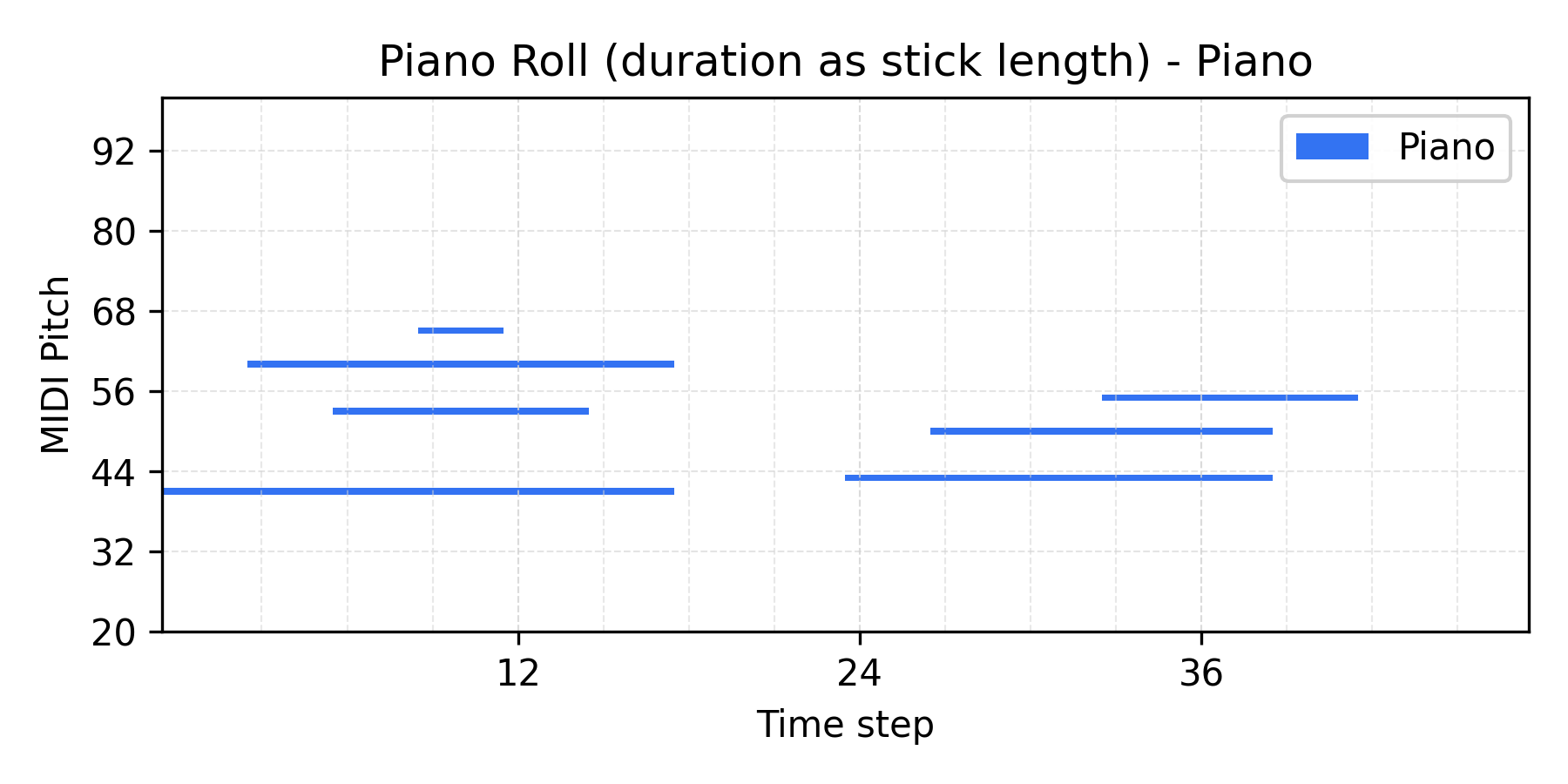}
            \caption{$\alpha = 0.4$}
        \end{subfigure}

        \vspace{0.5em}

        \begin{subfigure}{\textwidth}
            \includegraphics[width=\textwidth]{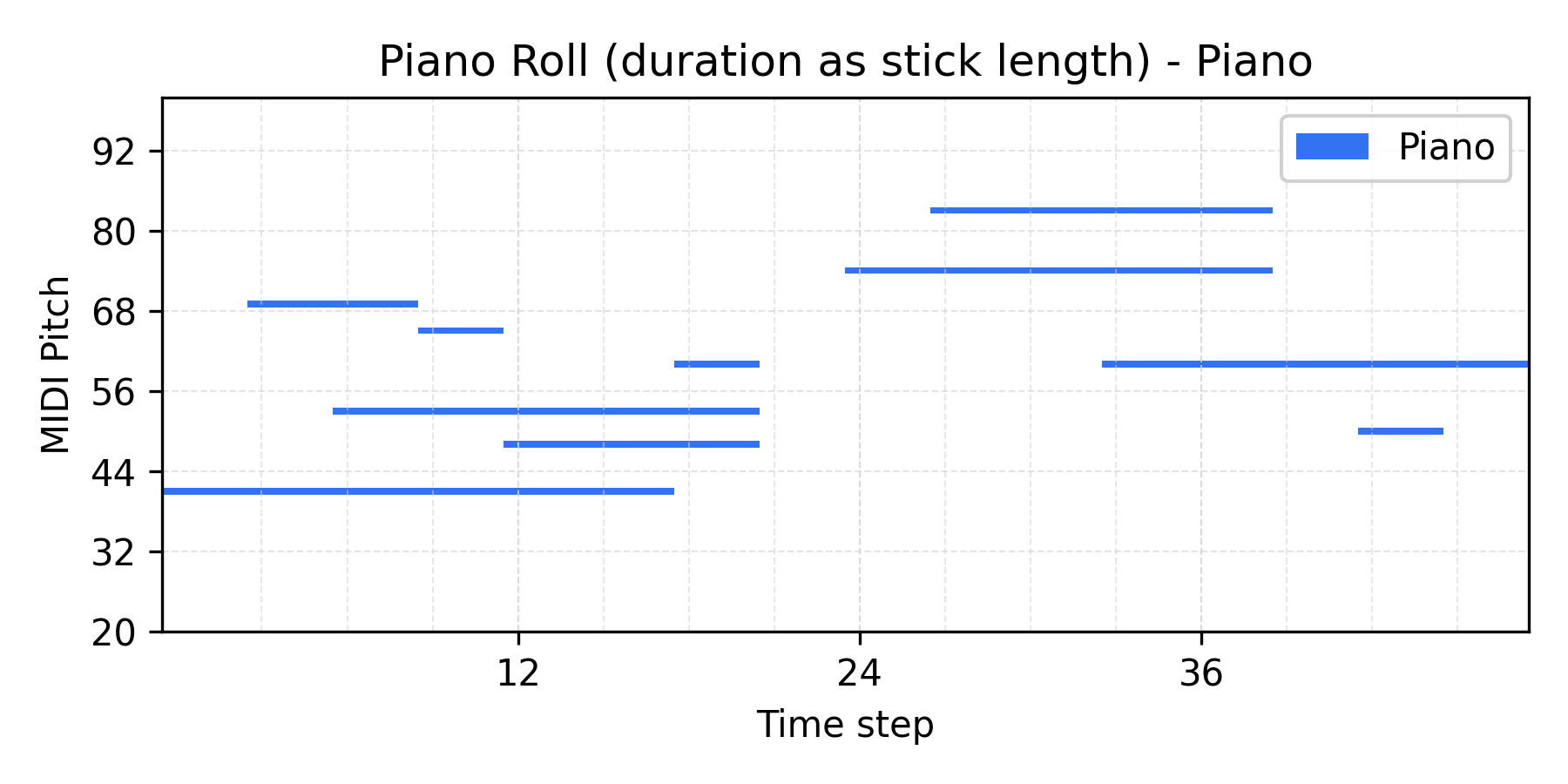}
            \caption{$\alpha = 0.5$}
        \end{subfigure}
    \end{minipage}
    \hfill
    \begin{minipage}{0.45\textwidth}
        \centering
        \begin{subfigure}{\textwidth}
            \includegraphics[width=\textwidth]{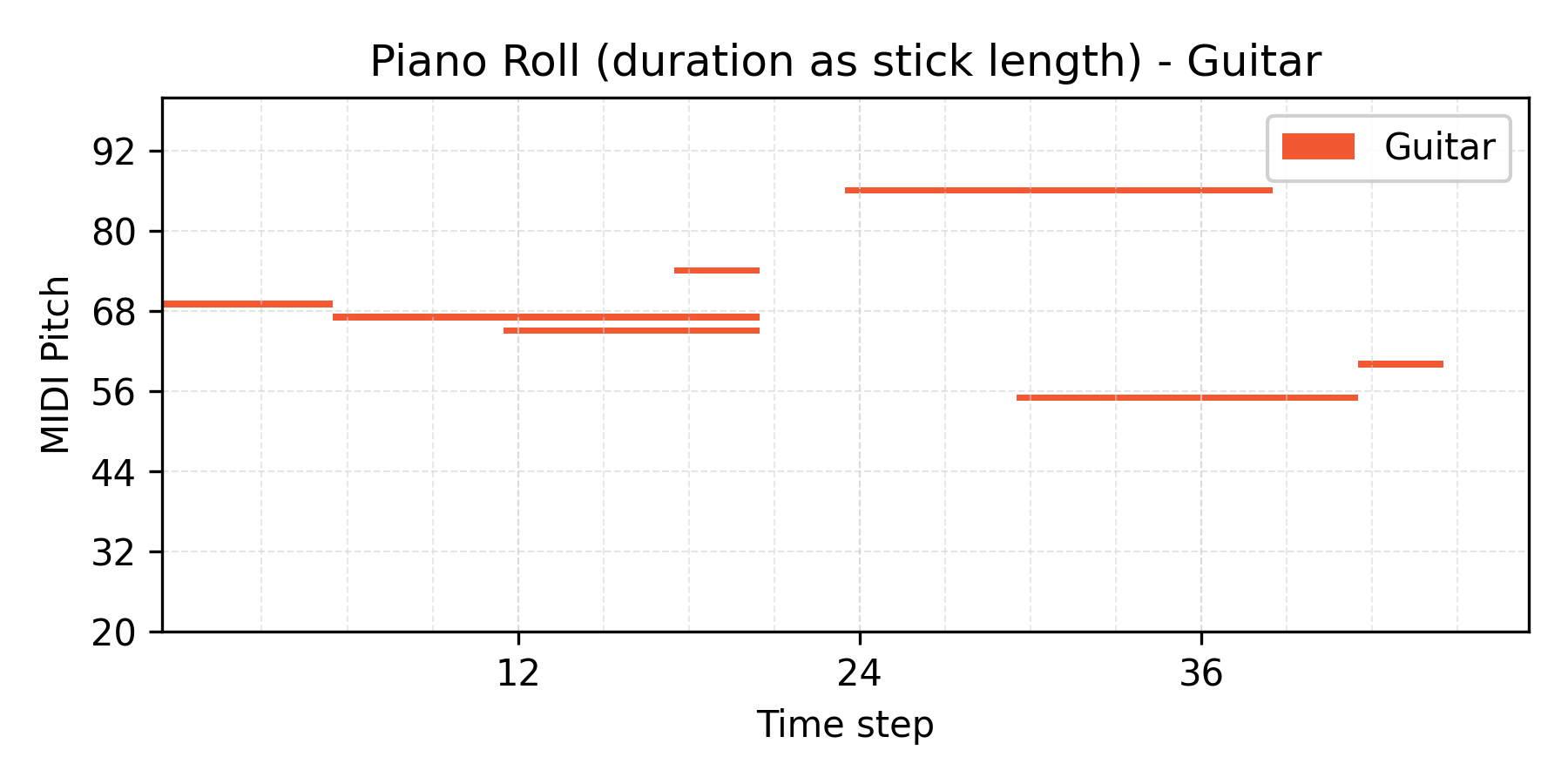}
            \caption{$\alpha = 0.6$}
        \end{subfigure}

        \vspace{0.5em}

        \begin{subfigure}{\textwidth}
            \includegraphics[width=\textwidth]{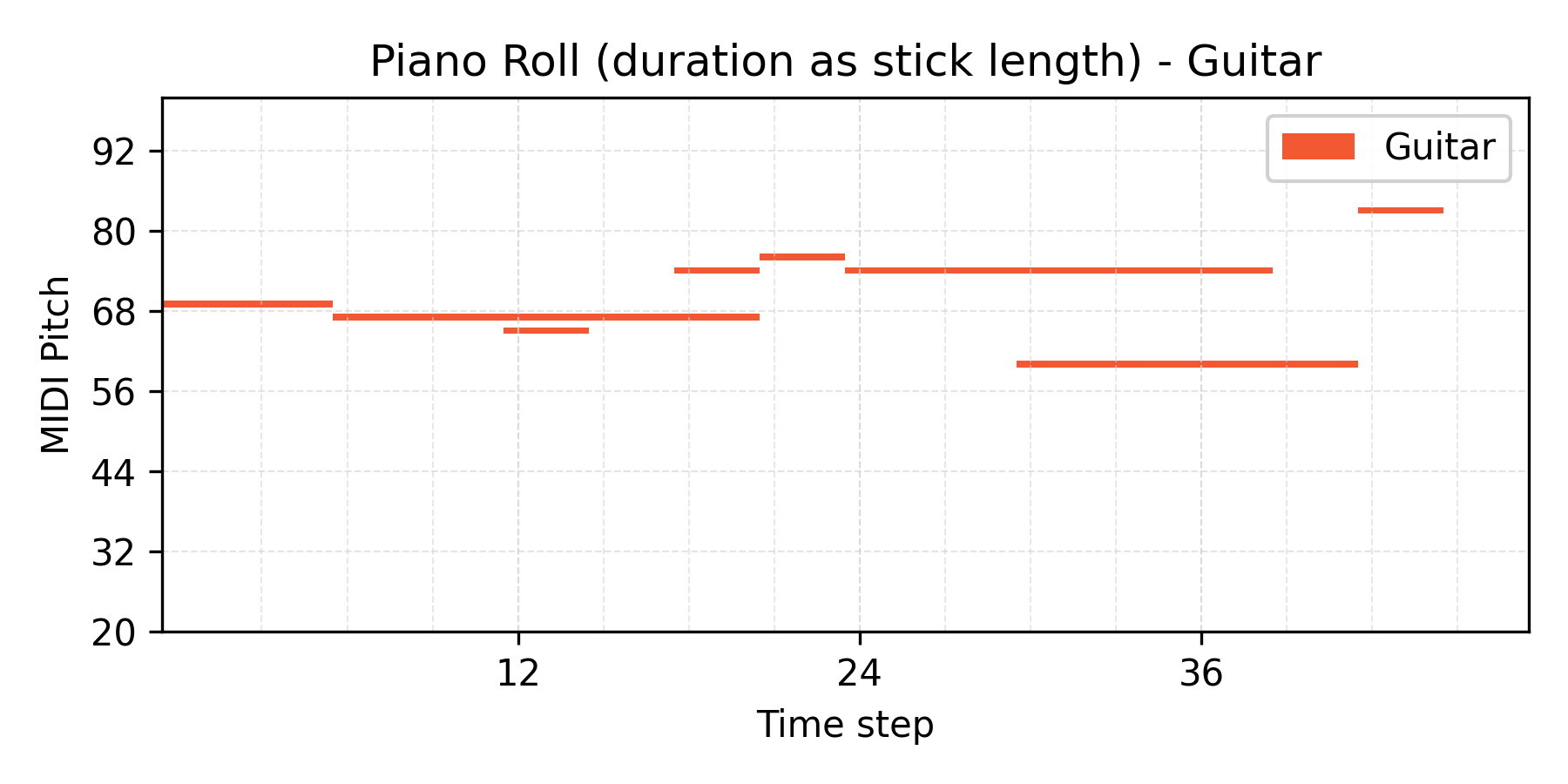}
            \caption{$\alpha = 0.7$}
        \end{subfigure}

        \vspace{0.5em}

        \begin{subfigure}{\textwidth}
            \includegraphics[width=\textwidth]{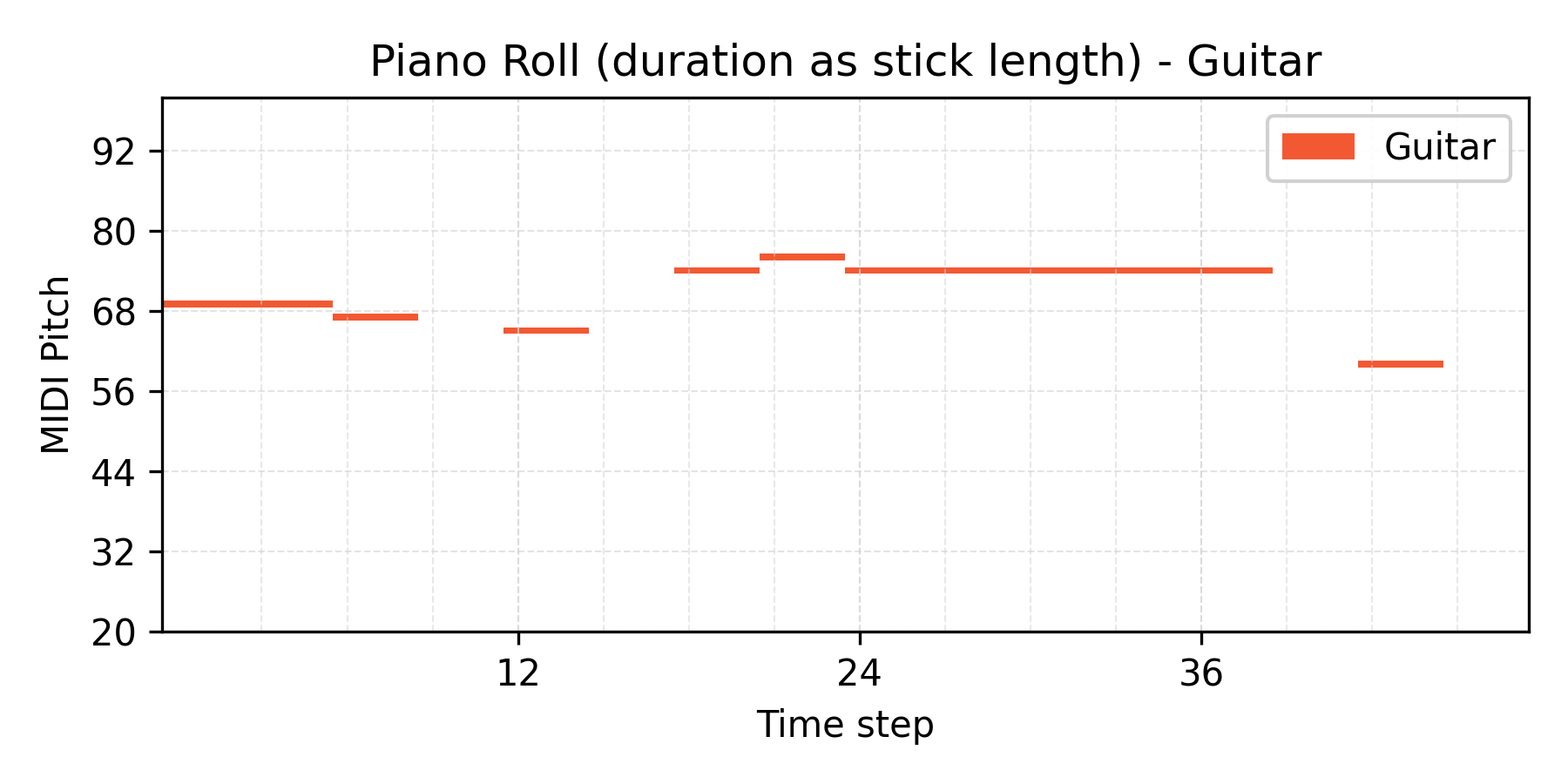}
            \caption{$\alpha = 0.8$}
        \end{subfigure}

        \vspace{0.5em}

        \begin{subfigure}{\textwidth}
            \includegraphics[width=\textwidth]{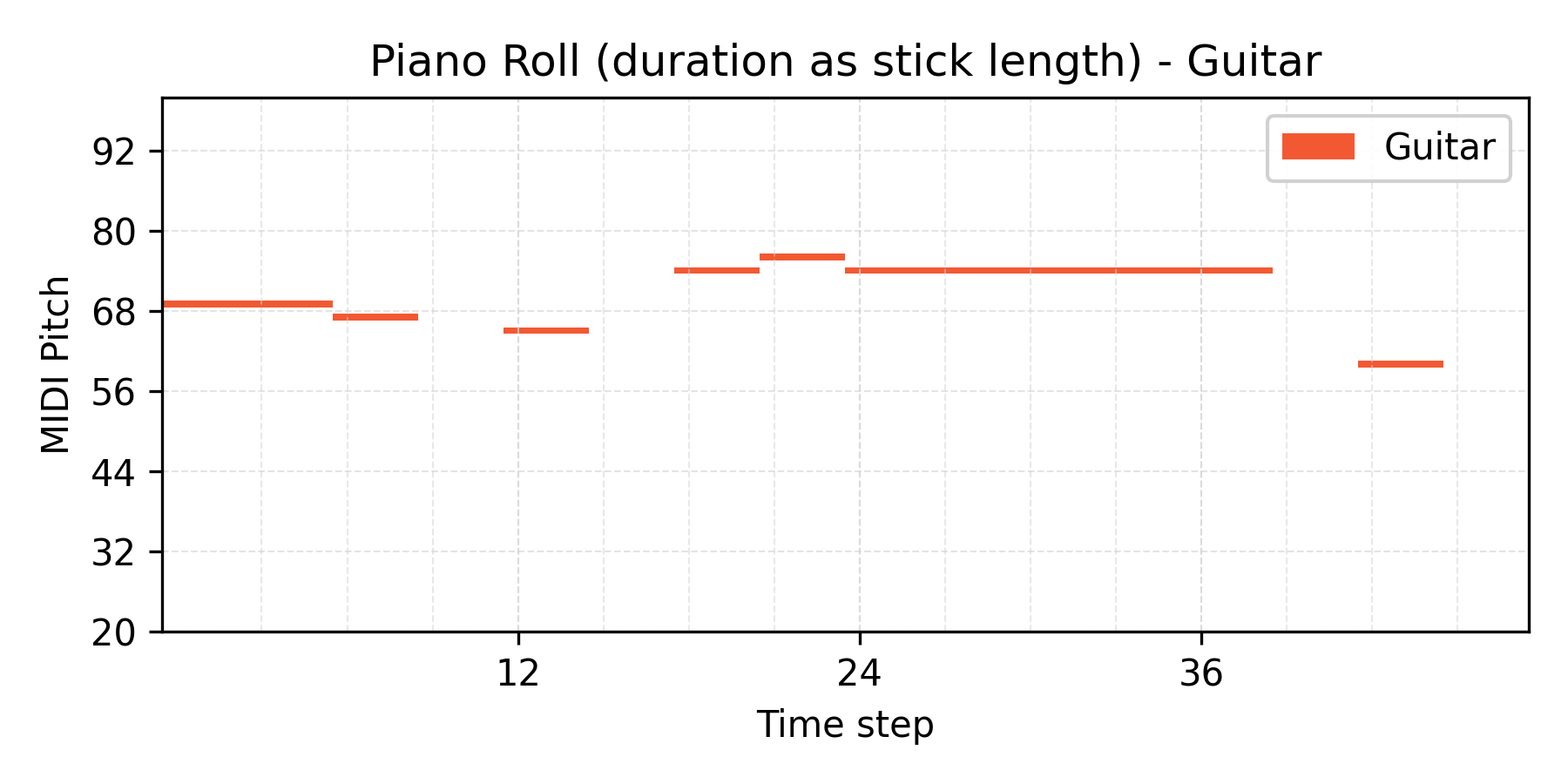}
            \caption{$\alpha = 1.0$}
        \end{subfigure}
    \end{minipage}

    \caption{Latent space interpolation results of PhraseVAE. Each subfigure shows the decoded phrase obtained by linearly interpolating between two phrase-level latent vectors with interpolation weight $\alpha$.}
    \label{fig:latent_interpolation}
\end{figure}

A desirable property of a variational latent representation is that semantic interpolation in latent space should correspond to smooth and meaningful transformations in the data domain. To examine this property, we perform linear interpolation between two randomly selected phrase-level latent vectors produced by PhraseVAE and decode the interpolated latents back into symbolic music.

Figure~\ref{fig:latent_interpolation} visualizes eight interpolation points with interpolation weights ranging from 0.0 to 1.0. The decoded results exhibit clear and consistent smoothness: musical attributes such as pitch, onset timing, duration, and instrument identity change gradually rather than abruptly. At each interpolation step, only a subset of note properties is modified, while the overall rhythmic contour and local textural patterns are largely preserved.

Notably, even when the interpolation traverses regions associated with different instruments, the decoded phrases remain musically coherent. Local note groupings and phrase-level “shapes” are often retained, suggesting that the latent space captures higher-level musical structure rather than surface-level token statistics. This behavior indicates that PhraseVAE learns a continuous and well-structured latent manifold aligned with musically meaningful dimensions.

Such smooth interpolation behavior is particularly important for downstream generative modeling. It implies that neighboring latent points correspond to perceptually similar musical phrases, reducing the risk of abrupt semantic jumps during diffusion-based generation. These results further support that PhraseVAE provides a suitable latent representation for long-range symbolic music modeling.

\subsection{Validation Loss}
Validation loss is a meaningful signal at the segment level (phrase or bar) and is used throughout all stages of PhraseVAE training.

However, validation loss at the \emph{song level} is far less informative. Melodies, harmonic progressions, and section layouts differ substantially across songs, so the distribution of phrase-level latents varies widely between the songs in the training set and that in the validaion set. During PhraseLDM training, we consistently observe that song-level validation loss increases steadily, yet this bears no correlation with generation quality. We therefore recommend using song-level validation loss sparingly.

\subsection{LDM Model Dimension}
Initially, we attempted to train the LDM using a model dimension equal to the VAE latent dimension. Under this configuration, the model was unable to fit even a single-song toy dataset. Increasing the \texttt{d\_model} size significantly alleviated this issue, whereas increasing depth alone did not help. For future work, we recommend setting the LDM model dimension to at least twice the latent dimensionality to ensure stable training.

\subsection{Other Evaluation Metrics}
Several other evaluation metrics exist but were not adopted here.

\textbf{FMD.}  
Proposed in prior work \cite{retkowski2024frechet} using the CLaMP-2 model \cite{wu2025clamp} as the feature extractor for FID-style evaluation. Since CLAMP-2 has a 32k-token context limit—sometimes still insufficient for full multitrack songs—FMD functions more as a segment-level metric. Given that our study focuses on local and full-song evaluation, we elected not to use it. Future work may incorporate FMD for broader comparison.

\textbf{Latent-Based Structure Metrics from \cite{wang2024whole}.}  
This metric evaluates intra-segment coherence and inter-segment distinction, useful for analyzing controllability under structure-conditioning. Because structure conditioning is not the central focus of this work, we did not include it, but it remains a strong candidate for future studies.

\subsection{Scalability of Structure Condition}
The current structural conditioning scheme is not scalable: it requires manually annotated section types and lengths for every song. While feasible for a dataset of roughly 1,000 songs, it is entirely impractical at the scale of millions or billions. Further, the concept of “section” itself—e.g., verse, chorus—varies widely across genres such as jazz, EDM, or ambient music. A rigid, universal taxonomy would introduce substantial noise.

To address these limitations, we outline two directions for future research.

\textbf{1. Extracting Section Layouts Directly from the SSM.}  
Self-Similarity Matrices can be computed automatically from MIDI or latent sequences and inherently encode sectional recurrence. Using SSM-based segmentation could bypass the need for human annotation.

\textbf{2. Replacing Section Labels with Automatically Extractable Musical Features.}  
We observe a strong correlation between instrumentation layout and sectional boundaries in POP909. For example, bar-to-bar instrumentation changes occur at section boundaries with probability \(59.2\%\), compared to \(33.8\%\) within the same section. When considering only instrument combinations (ignoring voice order), the difference becomes even more pronounced (\(49.2\%\) vs. \(22.9\%\)). This suggests that instrumentation implicitly encodes section-level structure.

We attempted to use instrumentation layout as a replacement structural condition. However, this produces very long conditioning sequences (e.g., four instrument tokens per bar; 400 tokens for a 100-bar song). On a small dataset of \(\sim800\) songs, this proved too difficult for the model to learn effectively. With substantially larger datasets, this direction may become viable.
\section{Summary}

This report presents a complete phrase-level latent diffusion framework for full-song symbolic music generation. By departing from the conventional note-attribute representation and instead modeling music at the phrase level, we transform the full-song generation problem from an extremely long-sequence modeling task into a compact latent modeling problem. PhraseVAE provides high-fidelity, well regularized phrase representations that preserve musical semantics while greatly reducing data dimensionality. PhraseLDM then operates entirely in this latent space, enabling efficient, non-autoregressive full-song generation with coherent local texture, idiomatic multitrack interaction, and meaningful global structure.

Through extensive analysis, we demonstrate that phrase-level latents strike a favorable balance between expressiveness, compactness, and trainability. The latent space supports diffusion modeling at a scale (128 bars per sequence) previously impractical for symbolic music. We further show that length conditioning and structure conditioning can reliably guide global form, while careful bottleneck design is essential for preventing memorization. Our evaluation also highlights the importance of new metrics—such as PhraseFID, SSM-based structural measures, and melody-level memorization tests—for assessing full-song symbolic generation models.

Overall, our results suggest that phrase-level latent diffusion is a promising and scalable paradigm for symbolic music generation. It removes the core bottlenecks associated with note-level modeling, enables fast and lightweight full-song generation, and produces musically convincing outputs even on modest datasets. We hope this work encourages future research to develop richer phrase representations, more scalable structural conditioning, and larger training corpora, ultimately advancing symbolic music generation toward more expressive, reliable, and globally coherent long-form composition.

\bibliography{_main}
\bibliographystyle{plain}

%%%%%%%%%%%%%%%%%%%%%%%%%%%%%%%%%%%%%%%%%%%%%%%%%%%%%%%%%%%%
% \newpage
% \appendix

% \input{appendices/1_remi_z}
% \input{appendices/2_implementation}
% \input{appendices/3_objective_metrics}
% \input{appendices/4_subjective_details}
% \input{appendices/5_probing_analysis}
% \input{appendices/6_broader_impact}
% \input{appendices/7_limitations}
%%%%%%%%%%%%%%%%%%%%%%%%%%%%%%%%%%%%%%%%%%%%%%%%%%%%%%%%%%%%

\newpage

\end{document}